\documentclass[11pt,a4paper]{amsart}
\usepackage{eurosym}
\usepackage{amsfonts}
\usepackage{hyperref}
\usepackage{amssymb}
\usepackage{amsmath}

\def\ii{{\,{\rm i}\,}}
\def\={\ =\ }

\def\e{{\,\rm e}\,}

\newcommand{\beq}{\begin{eqnarray}}
\newcommand{\eeq}{\end{eqnarray}}

\newcommand{\IC}{\mathbb{C}}

\theoremstyle{plain}

\numberwithin{equation}{section}

\begin{document}
\title[Matrix models in Donaldson-Thomas theory]{Matrix models and
stochastic growth in Donaldson-Thomas theory}
\date{July 2012 \hfill HWM--10--22 \ , \ EMPG--10--10}
\author{Richard J. Szabo}
\address{\flushleft Department of Mathematics, Heriot-Watt University, Colin
Maclaurin Building, Riccarton, Edinburgh EH14 4AS, UK, and Maxwell Institute
for Mathematical Sciences, Edinburgh, UK}
\email{R.J.Szabo@ma.hw.ac.uk}
\urladdr{}
\thanks{}
\author{Miguel Tierz}
\address{\flushleft Grupo de F\'{\i}sica Matem\'{a}tica, Complexo Interdisciplinar da Universidade de Lisboa, 
Av. Prof. Gama Pinto, 2, PT-1649-003 Lisboa, Portugal.}
\email{tierz@cii.fc.ul.pt}
\address{\flushleft Departamento de An\'{a}lisis Matem\'{a}tico, Facultad de Ciencias Matem\'{a}ticas. Universidad Complutense de Madrid. Plaza de Ciencias 3. 28040. Madrid. Spain.}
\email{tierz@mat.ucm.es}
\urladdr{}
\curraddr{ }

\begin{abstract}
We show that the partition functions which enumerate Donaldson-Thomas
invariants of local toric Calabi-Yau threefolds without compact divisors can
be expressed in terms of specializations of the Schur measure. We also
discuss the relevance of the Hall-Littlewood and Jack measures in the
context of BPS state counting and study the partition functions at arbitrary
points of the K\"ahler moduli space. This rewriting in terms of symmetric
functions leads to a unitary one-matrix model representation for
Donaldson-Thomas theory. We describe explicitly how this result is related
to the unitary matrix model description of Chern-Simons gauge theory. This
representation is used to show that the generating functions for
Donaldson-Thomas invariants are related to tau-functions of the integrable
Toda and Toeplitz lattice hierarchies. The matrix model also leads to an
interpretation of Donaldson-Thomas theory in terms of non-intersecting paths
in the lock-step model of vicious walkers. We further show that these
generating functions can be interpreted as normalization constants of a
corner growth/last-passage stochastic model.
\end{abstract}

\maketitle

\section{Introduction and summary of results}

Donaldson-Thomas theory computes enumerative invariants associated to the
number of points in the moduli spaces of ideal sheaves with trivial
determinant on a three-dimensional Calabi-Yau variety~\cite{DT,Thomas}. The
partition functions can be rephrased in terms of the counting of
noncommutative $U(1)$ instantons in a six-dimensional topological gauge
theory~\cite{INOV,CSS}. In this way the Donaldson-Thomas partition functions
may be regarded as generating functions which count BPS bound states of D0
and D2 branes in a single D6-brane, at least for appropriate values of the $%
B $-field. They can also be interpreted combinatorially in terms of the
enumeration of plane partitions (three-dimensional Young tableaux) with
boundary conditions along the three axes given by ordinary partitions (Young
diagrams), where the plane partitions are glued together along common
boundaries to form a crystal configuration. This leads to an interpretation
of Donaldson-Thomas theory in terms of the statistical mechanics of crystal
melting~\cite{ORV}. The melting crystal formulation connects
Donaldson-Thomas theory with topological string theory through the formalism
of the topological vertex~\cite{Aganagic:2003db}. For local toric
backgrounds, the generating functions for Donaldson-Thomas and Gromov-Witten
invariants are related by a simple change of
variables~\cite{MNOP,MOOP}.

As the physical moduli (e.g. the $B$-field) are continuously varied
this picture gets modified. Stable states may become unstable and
decay into more elementary constituents or new physical states can
appear in the spectrum. In Calabi-Yau compactifications it is only
for a special region of the moduli space that the stable objects are
enumerated via the Donaldson-Thomas invariants computed by
topological string theory. As one moves around the moduli space,
certain states can become lighter and different configurations become
energetically favoured over others. The moduli space can be divided
into chambers, each one with a physically distinct spectrum of stable
BPS states that depends on the value of the $B$-field through various two-cycles~\cite{Jafferis:2008uf,Chuang:2008aw,Aganagic:2009kf}. As the physical moduli are moved from one chamber to
another, crossing a so-called wall of marginal stability, the index
counting BPS states jumps according to a wall-crossing formula.

In
many cases one can solve for the
physical spectrum of BPS states. This is the case for the class of
examples of local toric threefolds without compact four-cycles where the
chamber structure of the moduli space has been explicitly constructed
and found a clear physical
interpretation via a lift to M-theory~\cite{Aganagic:2009kf}. Here
the partition function of BPS states at a generic point of the moduli
space is seen as receiving competing contributions from both M2-branes and
anti-M2-branes. In a certain region of the moduli space the anti-M2-brane states are all unstable and the partition function of BPS states
is purely holomorphic. This is the region around the large radius
point described by the topological string partition function $Z_{\rm
  top}(q,Q)$, with the parameter $q$ weighting D0-branes and the
parameters $Q$ weighting D2-branes. All the other regions can be
reached in principle by crossing walls of marginal stability and using
wall-crossing formulas. In another region of the moduli space the BPS state partition function has the form
\begin{equation}
Z_{\rm BPS} (q, Q) = Z_{\rm top} (q, Q) \ Z_{\rm top} (q, Q^{-1}) \ .
\end{equation}
This region corresponds to the noncommutative crepant resolution of a
toric singularity where the BPS states are computed by noncommutative
Donaldson-Thomas invariants. The counting of BPS states in this
region was introduced by Szendr\H{o}i for the conifold~\cite{Szendroi:2007nu}.

In this paper we will mostly work in the large radius and
noncommutative crepant resolution chambers of the K\"ahler moduli
space. We shall construct some new statistical mechanical models of
Donaldson-Thomas theory for local toric Calabi-Yau backgrounds which have no
compact divisors. We build upon the representation of these partition
functions in terms of random Young diagrams, which follows implicitly from
the original expansion of the topological vertex~\cite{Aganagic:2003db}. To
handle random Young diagrams, one needs to define a probability measure on
the set of Young tableaux. The classic example is the Plancherel measure,
introduced in the 1970's by Kerov and Vershik~\cite{Kerov}. It is of
interest in field theory and string theory, wherein the partition functions
of $\mathcal{N}=2$ supersymmetric gauge theory in four dimensions~\cite%
{NekOuk,LMN} and of topological string theory on local toric curves~\cite%
{Aganagic:2003db} can be conveniently expressed in terms of Plancherel and $%
q $-Plancherel measures.

More generally, one can define the Schur measure $\mathcal{M}_{\mathrm{Schur}%
}$, introduced by Okounkov \cite{O}, such that $\mathcal{M}_{\mathrm{Schur}%
}\{ \lambda \} $ for a partition $\lambda$ is proportional to $\mathfrak{s}%
_{\lambda }\left( x\right) \,\mathfrak{s}_{\lambda }\left( y\right) $. Here $%
\mathfrak{s}_{\lambda }(x)$ are the Schur polynomials, and $x$ and $y$ are
two independent (possibly infinite) sets of variables. There are several
important properties satisfied by the Schur measure.\footnote{%
See~\cite{O} for its more representation theoretic properties.} For
instance, as we shall see in Section~\ref{DTmeasure}, after proper
specification of the variables it contains the Plancherel and $q$-Plancherel
measures~\cite{Fulman,Strahov} as particular cases. Moreover, the Schur
measure has correlation functions of determinantal type, which is a common
feature of various problems in statistical mechanics, enumerative
combinatorics and probability theory that leads to their explicit solution.
One of the remarkable mathematical results of the last decade has been the
expression of probability measures on partitions in terms of determinantal
point processes \cite{Johanssonreview,review,Borodin}, which are very often
of random matrix type \cite{Johansson1,Johansson2}. Some straightforward
generalizations of the Schur measure that we shall discuss in Section~\ref%
{DTmeasure} are defined in terms of the Hall-Littlewood polynomials, or the
Jack polynomials which are one-parameter extensions of the Schur polynomials 
$\mathfrak{s}_{\lambda }\left( x\right) $~\cite{macdonald}.\footnote{%
The Macdonald polynomials \cite{macdonald} are the most general symmetric
polynomials known and they include all the other ones as special limiting
cases of their two parameters, but we shall not need them here.}

In this paper our interest in the Schur measure and its generalizations
rests in the property that the partition functions of Donaldson-Thomas
theory, and also of topological string theory and of the BPS state counting
for D6--D2--D0 branes~\cite{Szabo:2009vw}, on the backgrounds we consider
admit very natural expansions in terms of these measures. More precisely, as
we describe in detail in Section~\ref{DTmeasure}, they are normalization
constants of particular cases of such measures. These special instances come
from the specification of the two independent (infinite) sets of parameters $%
x$ and $y$ in terms of the variables of the generating functions for the
Donaldson-Thomas invariants~\cite{Szabo:2009vw}. Several mathematical
properties of these measures are utilized throughout this paper. Specific
examples of such rewritings in terms of symmetric functions can be found in 
\cite{Maeda:2004iq,Nakatsu:2007dk} in the context of the melting crystal
model with external potentials, and in \cite{Sulkowski:2008mx} in the
context of topological string theory (see also~\cite{Szabo:2009vw} for a
review).

This representation of the partition functions in terms of symmetric
functions is especially interesting due to the well-known combinatorics
theorem of Gessel \cite{Gessel}, which shows that the normalization constant
of the Schur measure can be written as a Toeplitz determinant. By exploiting
the classical Heine-Szeg\"{o} identity which links Toeplitz determinants
with unitary matrix models \cite{Szego}, it then leads to a unitary
one-matrix model representation for the Donaldson-Thomas partition
functions in the large radius and noncommutative crepant resolution chambers. This is the subject of Section~\ref{DTMM}. These matrix models
appear to be different from the ones previously found for Donaldson-Thomas
theory~\cite{Eynard:2009nd} and for topological string theory~\cite%
{Eynard:2010dh}, which involve infinite-dimensional hermitean multi-matrix
integrals with non-polynomial potentials. Since our unitary matrix model is
characterized by an infinite number of eigenvalues and its weight function
is a Jacobi theta-function, we can immediately relate it to the unitary
Chern-Simons matrix models~\cite{Okuda:2004mb,Tierz,Szabo:2010qv} with gauge
group $U(\infty )$. This is done by completing the unitary matrix model
representation of Chern-Simons theory by giving the partition function in
terms of a Toeplitz determinant.

The equivalent representations in terms of matrix models, Toeplitz
determinants and Schur measures imply, among other things, that the
generating functions for Donaldson-Thomas invariants are integrable. We
study this problem in detail in Section~\ref{DTint}. Using the work of Sato~%
\cite{Sato} and Segal-Wilson~\cite{Segal}, which show that tau-functions of
integrable hierarchies admit expansions in terms of Schur functions, we
relate the Donaldson-Thomas partition functions to tau-functions of the
integrable Toda lattice hierarchy. The equations of the hierarchy, together
with the string and divisor equations, uniquely determine the entire theory.
We consider various points of view on this issue in gauge theory and string
theory which suggest that, in appropriate instances, the pertinent
tau-function is either of 2-Toda or 1-Toda type. For example, in the free
fermion formulation~\cite{Maeda:2004iq,Nakatsu:2007dk} there is a hidden
symmetry yielding reduction to the 1-Toda lattice hierarchy, which is
related to the integrability structure of supersymmetric gauge theory in
four and five dimensions. On the other hand, the 1-Toda structure hidden in
the 2-Toda formalism is also evident in both ordinary and $q$-deformed
two-dimensional Yang-Mills theory. The results of~\cite{OkounToda} and~\cite[%
\S 1.4.2]{coverings} seem to hint that double Hurwitz numbers may be more
natural or better suited to describe the branched cover interpretation of
these gauge theories, which follows from the philosophy of our treatment as
well. This extends the usual description of two-dimensional Yang-Mills
theory in terms of simple Hurwitz numbers. Insofar as the $q$-deformed gauge
theory serves as a nonperturbative completion of the A-model topological
string theory on certain backgrounds, this integrability structure seems to
be related to the fact that equivariant Gromov-Witten theory can be either
of 1-Toda or of 2-Toda type. For example, topological string theory on the
resolved conifold belongs to the 1-Toda lattice hierarchy, while on toric
Fano surfaces it belongs to the 2-Toda lattice hierarchy~\cite{zhou}. The
latter geometries contain compact divisors and so fall out of the class of
backgrounds that our main line of development applies to in this paper. This
explanation for the absence of symmetry reduction of the Toda lattice
hierarchy appears to be a generic feature of topological string partition
functions.\footnote{%
This description appears to be equivalent to the topological sigma-model
description of Gromov-Witten theory on $\mathbb{P}^1$~\cite{Eguchi:1995er},
which is equivalent to a large $N$ hermitean one-matrix model with
non-polynomial potential that belongs to the 1-Toda lattice hierarchy. In
contrast, the equivariant Gromov-Witten theory of $\mathbb{P}^1$ is governed
by the 2-Toda lattice hierarchy~\cite{Okouneq}.}

A rather generic problem of determinantal type is that of non-intersecting
paths or non-intersecting Brownian motion in the continuum. The formalism
developed by Karlin and McGregor in the 1950's \cite{Karlin} provided
elegant determinantal expressions to describe non-intersecting Brownian
motion. These formulas were used in~\cite{dHT,dH} to express Wilson loop
observables in Chern-Simons gauge theory as probabilities in a model of $N$
non-intersecting Brownian motion particles, where $N$ is the rank of the
gauge group. The system of non-intersecting Brownian motion paths was
introduced by de Gennes to study chains of polymers under steric constraints~%
\cite{deGennes}. Later on, Fisher~\cite{fisher,fisheralso} introduced two
models of non-intersecting (vicious) random walkers in order to model domain
walls in two-dimensional lattice systems: the lock-step model and the
random-turns model. The latter model is intimately related to unitary random
matrices and the Gross-Witten model \cite{dHT}. It was also shown in \cite%
{dHT,dH} how some of the vicious walker expressions given by Fisher are
related to Chern-Simons observables.\ For example, the probability of
reunion for $N$ vicious walkers on a line gives the partition function of
Chern-Simons gauge theory on $S^{3}$ with gauge group $U(N)$. The problem
can be equivalently understood as the Brownian motion of a single particle
on the Weyl chamber of the gauge group. In \cite{dHT}, the eventual role of
the lock-step model in gauge theory was left as an open question.

The lock-step model of vicious walkers on a one-dimensional lattice allows
each walker at the tick of a clock to move either one lattice site to the
left or one lattice site to the right, with the restriction that no two
walkers may arrive at the same lattice site or pass one another. In Section~%
\ref{DTwalk} we will show that our matrix model expressions can be
interpreted in terms of non-intersecting paths in this model with infinitely
many vicious walkers. This follows directly from~\cite{lockstep} which gives
a vicious walker interpretation of the generic matrix averages that describe
the Donaldson-Thomas partition functions.

In addition, it turns out that the Schur measure employed here is the basis
of a generalized version of the corner growth model \cite%
{Johansson1,Johansson2}, and hence we can also interpret the
Donaldson-Thomas partition functions as normalization constants of this
stochastic process.\footnote{In~\cite{Orlando1,Orlando2} similar
  partition functions in two and three dimensions are related to 
growth processes described by the integrable XXZ spin chain and to a 
generalization thereof.} This model is intimately related to other random models,
such as the discrete polynuclear growth model, non-intersecting paths, and
random tilings of Aztec diamonds \cite{Johansson1,Johansson2} (see \cite%
{Johanssonreview} for a review). It is believed to belong to the
Kardar-Parisi-Zhang universality class for stochastic growth processes~\cite%
{KPZ}.

These two descriptions imply that Donaldson-Thomas theory has an
interpretation in terms of both discrete and continuous random systems. This
gives an indication that the non-intersecting paths system of \cite{lockstep}
is intimately related to the continuous last passage model in \cite%
{Johansson2}. Our treatment of the BPS bound state partition function holds
at all points in the K\"ahler moduli space, and hence naturally incorporates
its discontinuity due to wall-crossing phenomena. In these statistical
mechanics models, the jumps across walls of marginal stability have a
natural interpretation in terms of the creation or destruction of particles
and independent random variables. Furthermore, using results of~\cite%
{lockstep} we shall also see that the Donaldson-Thomas partition functions
are generating functions of certain types of infinite integer matrices that
satisfy specific symmetry conditions. It would be interesting to relate more
directly this enumerative interpretation to the actual integer
Donaldson-Thomas invariants that count ideal sheaves (equivalently
instantons or D6--D2--D0 bound states).

\subsection*{Note added}

While we were completing the present paper, the preprint~\cite{Ooguri:2010yk}
appeared, whose results overlap with ours, but with a different approach and
theme. Their derivation of the unitary matrix models is based on observing
that the Chern-Simons matrix model is related to the MacMahon function by
expansion of the weight function of the matrix model. Based on the melting
crystal formalism, both a free fermion formulation and the Gessel-Viennot
determinantal expression which counts non-intersecting paths is then
employed to find explicit expressions for the matrix models. As summarised
above, our approach is different and yields these same results within a
different setting. For example, the non-uniqueness of the weight function of
the matrix model, discussed after eq.~(4.20) in~\cite{Ooguri:2010yk}, is
emphasised below in (\ref{mat1}) and the discussion afterwards in a
completely different way; this has also been noticed previously for
Chern-Simons matrix models. Notice that the Lindstr\"{o}m-Gessel-Viennot
theorem used in~\cite{Ooguri:2010yk} is essentially the same as the
Karlin-McGregor theorem that was used previously in~\cite{dHT,dH} to obtain
Chern-Simons observables. Although the Karlin-McGregor theorem leads to a
two-matrix model, it is explained in~\cite{dH} how to relate it to the
one-matrix model of Chern-Simons gauge theory. This result is also used by
Johansson to find the Schur measure (\ref{G(M,N)}) in the study of the
generalized corner growth model that we employ in Section~\ref{DTwalk}.

\subsection*{Acknowledgments}

This work was supported in part by grant ST/G000514/1 \textquotedblleft
String Theory Scotland\textquotedblright\ from the UK Science and Technology
Facilities Council. The work of MT has been partially supported by the project 
"Probabilistic approach to finite and infinite dimensional dynamical systems" 
(PTDC/MAT/104173/2008) at the Universidade de Lisboa.

\section{Donaldson-Thomas theory and symmetric function measures on
partitions\label{DTmeasure}}

\subsection{Affine space}

The partition function of Donaldson-Thomas theory in the simplest case of
the Calabi-Yau threefold $%
\mathbb{C}
^{3}$ is given by the MacMahon function%
\begin{equation}
Z_{\mathrm{DT}}^{%
\mathbb{C}
^{3}}\left( q\right) =\prod\limits_{n=1}^{\infty }\, \frac{1}{\left(
1-q^{n}\right) ^{n}}\equiv M(q) \ ,  \label{DT}
\end{equation}%
where we choose a string coupling constant $g_s$ such that the quantum
parameter $q\equiv {\,\mathrm{e}}\,^{-g_s}$ satisfies $|q|<1$. This is the
generating function for plane partitions \cite{MacMahon,Stanley}. They
correspond to pointlike instantons, or D0-branes, in the D6-brane gauge
theory on $\mathbb{C}^3$, with charge equal to the number of boxes in the
plane partition. Each plane partition is equivalent to a monomial ideal
which corresponds to an ideal sheaf on $\mathbb{C}^3$. On the other
hand, the topological string partition function in this case is
$Z_{\rm top}^{\mathbb{C}^3}(q)= M(q)^{1/2}$; in general the relation between the two
generating functions in the large radius chamber is given by~\cite{MNOP,MOOP}
\beq
Z^X_{\rm
  DT}(q,Q)= M(q)^{\chi(X)/2}\, Z_{\rm top}^X(q,Q) \ ,
\label{ZDTZtoprel}\eeq
where $\chi(X)$ is the topological Euler character of $X$; in the
present case we use the convention $\chi(\IC^3)=1$.

As pointed out by~\cite{ORV,Maeda:2004iq,Nakatsu:2007dk,Szabo:2009vw}, this
partition function can be rewritten in terms of Schur functions. This
expression is the crux of the equivalence between Donaldson-Thomas and
Gromov-Witten theories in the toric case~\cite{MNOP,MOOP,INOV} within the melting crystal formulation~\cite%
{ORV}. However, it also follows directly from the enumerative expression.
The Schur polynomials $\mathfrak{s}_{\lambda }(x)$ \cite{Stanley,macdonald}
constitute a basis of symmetric functions in a given set of variables $%
x=(x_{i})_{i\geq1}$ and are indexed by Young diagrams (ordinary partitions) $%
\lambda=(\lambda_i)_{i\geq1} $, with $\lambda_i\geq\lambda_{i+1}\geq0$
giving the length of the $i$-th row. If the variables $x$ are regarded as
eigenvalues of some matrix $M\in sl_{n}$, then $\mathfrak{s}_{\lambda
}(x)\equiv \mathrm{Tr}_{\lambda }(M)$ is the trace of $M$ in the irreducible 
$sl_n$-representation associated to $\lambda $. The Schur polynomials may
also be more explicitly defined in terms of the skew-symmetric polynomials $%
\mathfrak{a}_{\mu }=\det_{i,j} (x_{i}^{\mu _{j}+n-j})$ as $\mathfrak{s}%
_{\lambda }(x)\equiv \mathfrak{a}_{\lambda}(x)/\mathfrak{a}_{0}(x)$. By
using the Cauchy identity~\cite{Stanley}%
\begin{equation}
\sum_{\lambda }\, \mathfrak{s}_{\lambda }(x)\, \mathfrak{s}_{\lambda
}(y)=\prod\limits_{i,j\geq 1}\, \frac{1}{1-x_{i}\,y_{j}}\equiv \mathcal{Z} \
,  \label{SC}
\end{equation}%
and considering the case of an infinite number of variables with the
specializations $x_{i}=q^{i-1/2}$ and $y_{j}=q^{j-1/2},$ then $\left( \ref%
{SC}\right) $ directly gives the expression \cite{Szabo:2009vw} for the
Donaldson-Thomas partition function of $%
\mathbb{C}
^{3}$ in terms of Schur functions, 
\begin{equation}
Z_{\mathrm{DT}}^{%
\mathbb{C}
^{3}}\left( q\right) =\sum_{\lambda }\, \mathfrak{s}_{\lambda }\big(q^{{i-1}/%
{2}}\big)^{2} \ .  \label{DTc}
\end{equation}

Let us phrase this result in terms of the Schur measure introduced by
Okounkov in \cite{O}. It assigns to each partition $\lambda $ the weight $%
\mathcal{M}_{\mathrm{Schur}}\{\lambda\}=\frac1{\mathcal{Z}}\, \mathfrak{s}%
_{\lambda }(x)\,\mathfrak{s}_{\lambda }(y),$ where $\mathfrak{s}_{\lambda
}(x)$ are Schur functions. Then%
\begin{equation}
\mathcal{P}_N\{ \lambda\} =\frac1{\mathcal{Z}}~ \sum_{\lambda \, :\,\lambda
_{1}\leq N}\, \mathfrak{s}_{\lambda }(x)\, \mathfrak{s}_{\lambda }(y)
\end{equation}%
is the probability that the number of boxes in the first row of the
associated Young diagram is $\leq N$. Taking the limit $N\rightarrow \infty $%
, the normalization constant $\mathcal{Z}$ is given by (\ref{SC}). Thus the
Donaldson-Thomas partition function $\left( \ref{DTc}\right) $ is given by
the normalization constant of the Schur measure, specialized at $%
x_{i}=q^{i-1/2}$ and $y_{j}=q^{j-1/2}$.

By making the specification $\mathfrak{s}_{\lambda }(1,1,\dots)= \dim
\lambda $, the Schur measure contains the Plancherel measure%
\begin{equation}
\mathcal{M}_{\mathrm{Planch}}\{\lambda\} =\Big(\,\frac{\dim\lambda}{|\lambda|\,
!}\, \Big)^2 \ ,  \label{Plancherel}
\end{equation}%
where the dimension of the corresponding irreducible representation of the
symmetric group $S_{|\lambda|}$, with $|\lambda|:=\sum_i\, \lambda_i$ the
weight of the representation, is given by 
\begin{equation}
\dim\lambda=\prod\limits_{u\in \lambda }\, \frac{1}{{h}(u)} \ .
\end{equation}%
Here ${h}(u)\equiv \lambda _{i}+\lambda _{j}^{\prime }-i-j+1 $ is the hook
length of the box $u=(i,j)$ in $\lambda$, and $\lambda ^{\prime }$ denotes
the conjugate partition to $\lambda$.\footnote{%
The conjugate of a partition is obtained by interchanging rows and columns
in its Ferrers graph.} Our specification is instead $q^{-{\left\vert \lambda
\right\vert }/{2}}\, \mathfrak{s}_{\lambda }(1,q,\dots,q^{n-1}),$ and we
know that \cite{Stanley,macdonald}%
\begin{equation}
\mathfrak{s}_{\lambda }(1,q,\dots,q^{n-1})=q^{n(\lambda )}\,
\prod\limits_{u\in \lambda }\, \frac{\left[ n+c\left( u\right) \right] }{%
\left[ h\left( u\right) \right] } \ ,  \label{qdim}
\end{equation}%
where $n(\lambda )\equiv \sum_{i\geq 1}\, (i-1)\, \lambda _{i}$ and, for
each box $u=(i,j)$ of the diagram $\lambda$, $c(u)\equiv j-i$ is the content
of $u.$ The square brackets here denote $q$-numbers, 
\begin{eqnarray}
[a]\equiv \frac{q^{a/2}-q^{-a/2}}{q^{1/2}-q^{-1/2}} \ ,
\end{eqnarray}
and the right-hand side of $\left( \ref{qdim}\right) $ is the $q$%
-deformation of the dimensions of $sl_n$ representations, i.e. the quantum
dimensions $\dim_q\lambda$~\cite{fuchs}. Then the Schur measure with this
specialization is a $q$-Plancherel measure, essentially the $q$-deformed
Plancherel measure discussed in \cite{Fulman}.\footnote{%
Other $q$-deformations are introduced in \cite{Strahov}, where only one of
the Schur functions is $q$-specialized.} Thus the normalization constant of
the $q$-Plancherel measure gives the Donaldson-Thomas partition function on $%
\mathbb{C}
^{3}$.

\subsection{Conifold}

There are two non-isomorphic crepant resolutions of the conifold singularity 
$z_{1}\,z_{2}-z_{3}\,z_{4}=0$ in
$\mathbb{C}^4$~\cite{Szendroi:2007nu}. Using (\ref{ZDTZtoprel}) with $\chi(X)=4$, the
topological string partition function of the resolved conifold $\mathbb{X}=%
\mathcal{O}_{\mathbb{P}^1}\left( -1)\oplus\mathcal{O}_{\mathbb{P}%
^1}(-1\right) $ can be expressed in terms of the Donaldson-Thomas generating
function 
\begin{equation}
Z_{\mathrm{DT}}^{\mathbb{X}}\left( q,Q\right) = M(-q)^{2}\, M(Q,-q)^{-1}
\label{SzX}
\end{equation}
where $Q={\,\mathrm{e}}\,^{-t}$ with $t$ the K\"ahler modulus of the base $%
\mathbb{P}^1\hookrightarrow \mathbb{X}$ such that $|Q|<1$, and 
\begin{equation}
M(Q,q)= \prod_{n=1}^\infty\, \frac1{\left(1-Q\,q^n\right)^n}
\end{equation}
is the generalized MacMahon function, the generating function for weighted
plane partitions. The extra parameter $Q$ weights the contributions from
``fractional instantons'' stuck at the resolution of the conifold
singularity, or D2-branes in the D6-brane gauge theory on $\mathbb{X}$, with
charge equal to the number of diagonal boxes of the plane
partition. From the perspective of the chamber analysis
of~\cite{Szendroi:2007nu,Jafferis:2008uf}, the partition function
(\ref{SzX}) gives the BPS state counting in the large radius chamber
of the K\"ahler moduli space. On the
other hand, the Donaldson-Thomas theory of the noncommutative crepant
resolution of the conifold singularity enumerates framed cyclic
representations of a quiver with the Klebanov-Witten superpotential~\cite%
{Szendroi:2007nu}, or equivalently of the corresponding quotient path
algebra $\mathbb{A}$. The corresponding partition function can be computed
by counting partitions of a two-coloured pyramid of length one,\footnote{%
The two colours are weighted by the variables $q_0=q/Q$ and $q_1=Q$.}
associated with a perfect matching of a brane tiling, and is given by \cite%
{Szendroi:2007nu} 
\begin{equation}
Z_{\mathrm{DT}}^{\mathbb{A}}\left( q,Q\right) =M(-q)^{2}\, M(Q,-q)^{-1}\,
M(Q^{-1},-q)^{-1} \ .  \label{Sz}
\end{equation}

The appropriate measure in this case involves the Hall-Littlewood
polynomials $\mathfrak{p}_{\lambda }(x;v)$ \cite{macdonald}, one of two
generalizations of the Schur polynomials. They are symmetric polynomials in $%
x,$ homogeneous of degree $\left\vert \lambda \right\vert$, and with
coefficients in $%
\mathbb{Z}
\left[v\right] $. As with the Schur polynomials, the Hall-Littlewood
polynomials can be extended to Hall-Littlewood functions involving an
infinite number of variables. They are defined by%
\begin{eqnarray}
\mathfrak{p}_{\lambda }(x;v) =\sum_{\sigma \in S_{n}/S_{n}^{\lambda }}\,
\sigma \Big( x\, \prod\limits_{\lambda _{i}>\lambda _{j}}\, \frac{x_{i}-v\,
x_{j}}{x_{i}-x_{j}}\Big) \ ,
\end{eqnarray}%
where $S_{n}^{\lambda }$ is the subgroup of the symmetric group $S_{n}$
consisting of permutations that leave $\lambda=(\lambda_1,\dots,\lambda_n) $
invariant, and $\sigma \left( f(x)\right) =f\left( \sigma \left( x\right)
\right) $ with $\sigma(x)=x_1^{\sigma_1}\cdots x_n^{\sigma_n}$. The
parameter $v$ serves to interpolate between the Schur polynomials $\mathfrak{%
s}_{\lambda }(x)$ at $v=0$ and the monomial symmetric functions $\mathfrak{m}%
_{\lambda }(x)=\sum_{\sigma\in S_n/S_n^\lambda}\,\sigma(x)$ at $v=1$. The
corresponding Cauchy identity for Hall-Littlewood polynomials is%
\begin{equation}
\sum_{\lambda }\, b_\lambda(v)\, \mathfrak{p}_{\lambda }(x;v)\, \mathfrak{p}%
_{\lambda }(y;v)=\prod\limits_{i,j\geq 1}\, \frac{1-v\, x_{i}\, y_{j}}{%
1-x_{i}\, y_{j}} \equiv \mathcal{Z}_{\mathrm{HL}} \ ,  \label{PC}
\end{equation}%
where 
\begin{equation}
b_{\lambda }\left(v\right) =\prod\limits_{i=1}^{\lambda _{1}}\,
(v)_{m_{i}\left( \lambda \right) } \ .
\end{equation}
Here $(v)_0\equiv 1$, $(v)_m\equiv \prod_{1\leq j\leq m}\, \big(1-v^j\big)$
for $m\in\mathbb{N}$, and $(v)_\infty\equiv \prod_{j\in\mathbb{N}}\, \big(%
1-v^j\big)$, while $m_i(\lambda)$ is the number of parts of $\lambda$ equal
to $i$.

In parallel to the $%
\mathbb{C}
^{3}$ case above, if we make the specification $x_{i}=(-q)^{i-1/2}$ and $%
y_{j}=(-q)^{j-1/2}$ in (\ref{PC}), and set the parameter $v$ equal to either 
$Q$ or $Q^{-1},$ then we obtain expressions for the conifold
Donaldson-Thomas partition functions analogous to $\left( \ref{DTc}\right)$
given by 
\begin{equation}
Z_{\mathrm{DT}}^{\mathbb{X}}\left( q,Q\right) =M(-q)\, \sum_{\lambda }\,
b_{\lambda }\left(Q\right)\, \mathfrak{p}_{\lambda }\big(\left( -q\right)
^{i-{1}/{2}};Q \big)^{2}  \label{X}
\end{equation}%
and 
\begin{equation}
Z_{\mathrm{DT}}^{\mathbb{A}}\left( q,Q\right) =\sum_{\lambda }\, b_{\lambda
}\left(Q\right) \, \mathfrak{p}_{\lambda }\big(\left( -q\right) ^{i-{1}/{2}%
};Q \big)^{2}~ \sum_{\mu }\, b_{\mu }\left(Q^{-1}\right)\, \mathfrak{p}_{\mu
}\big(\left( -q\right) ^{i-{1}/{2}};Q^{-1} \big)^{2} \ .  \label{A}
\end{equation}%
Likewise, one can define a Hall-Littlewood measure on partitions given by%
\begin{equation}
\mathcal{M}_{\mathrm{HL}}\left\{ \lambda \right\} =\frac{1}{\mathcal{Z}_{%
\mathrm{HL}}}\, b_\lambda(v)\, \mathfrak{p}_{\lambda }(x;v)\, \mathfrak{p}%
_{\lambda }(y;v) \ ,
\end{equation}%
with the normalization $\mathcal{Z}_{\mathrm{HL}}$ given by $\left( \ref{PC}%
\right)$. This leads to the Donaldson-Thomas partition functions on the
conifold $\left( \ref{X}\right) $ and $\left( \ref{A}\right) $, when
specializing to $x_{i}=(-q)^{i-1/2} $ and $y_{j}=(-q)^{j-1/2}$. This
specialized measure can be regarded as a Hall-Littlewood deformation of the $%
q$-Plancherel measure.

We have seen that the generalization of symmetric functions provided by the
Hall-Littlewood polynomials is appropriate to the Donaldson-Thomas partition
functions of the conifold, in contrast to the simpler case of $%
\mathbb{C}
^{3}$. This is especially true in the noncommutative setting $\left( \ref{A}%
\right) $, where even the two MacMahon function factors in (\ref{Sz}) are
automatically included. The appearance of Hall-Littlewood polynomials in
topological string theory has already been discussed in~\cite%
{Sulkowski:2008mx}. In particular, the expression (\ref{X}) for the
partition function of the resolved conifold is given there. However, the
reduced partition function 
\begin{equation}
\widetilde{Z}_{\mathrm{DT}}^{\,{\mathbb{X}}}\left( q,Q\right) = \frac{Z_{%
\mathrm{DT}}^{\mathbb{X}}\left( q,Q\right) }{M(-q)^{2}}=M(Q,-q)^{-1}
\end{equation}%
factors out the contributions from the degree zero subschemes (regular
D0-branes) of $\mathbb{X}$, and it can be written solely in terms of Schur
functions by using the dual Cauchy identity%
\begin{equation}
\sum_{\lambda }\, \mathfrak{s}_{\lambda }(x)\, \mathfrak{s}_{\lambda
^{\prime }}(y)=\prod\limits_{i,j\geq 1}\, \left( 1+x_{i}\, y_{j}\right)
\label{dualC}
\end{equation}%
together with the scaling property $\mathfrak{s}_{\lambda }(Q\,
x)=Q^{|\lambda|}\, \mathfrak{s}_{\lambda }(x)$ to write 
\begin{equation}
\widetilde{Z}_{\mathrm{DT}}^{\,{\mathbb{X}}}\left( q,Q\right) =\sum_{\lambda
}\, (-Q)^{\left\vert \lambda \right\vert }\, \mathfrak{s}_{\lambda }\big(%
(-q)^{i-1/2}\big)\, \mathfrak{s}_{\lambda ^{\prime }}\big((-q)^{i-1/2}\big) %
\ .  \label{dual}
\end{equation}%
Similarly, one has 
\begin{eqnarray}
&&  \label{ZA} \\
&& \widetilde{Z}_{\mathrm{DT}}^{\,{\mathbb{A}}}\left( q,Q\right)=
\sum_{\lambda,\mu }\, (-Q)^{\left\vert \lambda \right\vert-\left\vert
\mu\right\vert }\, \mathfrak{s}_{\lambda }\big((-q)^{i-1/2}\big)\, \mathfrak{%
s}_{\lambda ^{\prime }}\big((-q)^{i-1/2}\big) \, \mathfrak{s}_{\mu}\big(%
(-q)^{i-1/2}\big)\, \mathfrak{s}_{\mu^{\prime }}\big((-q)^{i-1/2}\big) \ . 
\notag
\end{eqnarray}

\subsection{Orbifold}

Consider the action of the cyclic group $\mathbb{Z}_2$ on $\mathbb{C}^3$
generated by $(z_1,z_2,z_3)\mapsto (-z_1,-z_2,z_3)$. The crepant resolution
of the orbifold singularity $\mathbb{C}^3/\mathbb{Z}_2$ given by the $%
\mathbb{Z}_2$-Hilbert scheme is ${\mathbb{Y}}=\mathcal{O}_{\mathbb{P}%
^1}(-2)\oplus\mathcal{O}_{\mathbb{P}^1}(0)= \mathcal{O}_{\mathbb{P}%
^1}(-2)\times\mathbb{C}$, where the first factor is the minimal
(Hirzebruch-Jung) resolution of the $A_1$ Klein singularity $\mathbb{C}^2/%
\mathbb{Z}_2$~\cite{Szabo:2009vw}. The corresponding Donaldson-Thomas
partition function is similar to that of the conifold (\ref{SzX}) and reads~%
\cite{Young} 
\begin{equation}
Z_{\mathrm{DT}}^{\mathbb{Y}}(q,Q)=M(-q)^2\, M(Q,-q) \ ,  \label{OrbY}
\end{equation}
where again $Q={\,\mathrm{e}}\,^{-t}$. The Donaldson-Thomas invariants of
the noncommutative crepant resolution, given by the quiver algebra of the
McKay quiver associated to the $A_1$ singularity~\cite{Szabo:2009vw}, are
the same as the orbifold Donaldson-Thomas invariants of the quotient stack $[%
\mathbb{C}^3/\mathbb{Z}_2]$~\cite{Young,JoyceSong}. They are labelled by $%
\mathbb{Z}_2$-representations $\rho$ and enumerate $\mathbb{Z}_2$-invariant
zero-dimensional subschemes $Y\subset\mathbb{C}^3$ with $H^0(\mathcal{O}%
_Y)=\rho$. The corresponding partition function can be computed by counting
configurations of two-coloured boxes, with the colours corresponding to the
two irreducible representations of $\mathbb{Z}_2$, and is given by~\cite%
{Young} 
\begin{equation}
Z_{\mathrm{DT}}^{\mathbb{C}^3/\mathbb{Z}_2}(q,Q)=M(-q)^2\, M(Q,-q)\,
M(Q^{-1},-q) \ .  \label{Orb}
\end{equation}

The reduced orbifold partition functions are ``dual'' to those of the
conifold, in the sense that they are related through 
\begin{eqnarray}
\widetilde{Z}_{\mathrm{DT}}^{\,{\mathbb{Y}}}(q,Q)=\widetilde{Z}_{\mathrm{DT}%
}^{\,\mathbb{X}}(q,Q)^{-1} \qquad \mbox{and} \qquad \widetilde{Z}_{\mathrm{DT%
}}^{\,\mathbb{C}^3/\mathbb{Z}_2}(q,Q) =\widetilde{Z}_{\mathrm{DT}}^{\,%
\mathbb{A}}(q,Q)^{-1} \ .  \label{DTdual}
\end{eqnarray}
Hence one can regard the orbifold generating functions in terms of
Hall-Littlewood measures by using the formulas of the preceding subsection.
Alternatively, they can also be written solely in terms of Schur functions
as 
\begin{eqnarray}
\widetilde{Z}_{\mathrm{DT}}^{\,{\mathbb{Y}}}\left( q,Q\right)=\sum_{\lambda
}\, Q^{\left\vert \lambda \right\vert }\,\mathfrak{s}_{\lambda }\big(%
(-q)^{i-1/2}\big)^{2}  \label{inv}
\end{eqnarray}
and 
\begin{eqnarray}
\widetilde{Z}_{\mathrm{DT}}^{\,\mathbb{C}^3/\mathbb{Z}_2}\left(
q,Q\right)=\sum_{\lambda,\mu }\, Q^{\left\vert \lambda \right\vert -|\mu|}\,%
\mathfrak{s}_{\lambda }\big((-q)^{i-1/2}\big)^{2}\, \mathfrak{s}_{\mu }\big(%
(-q)^{i-1/2}\big)^{2} \ .  \label{invorb}
\end{eqnarray}

\subsection{BPS state counting}

For the toric Calabi-Yau backgrounds $X$ of interest in this paper, which
have no compact divisors, it was shown in~\cite%
{Aganagic:2009kf,Cecotti:2009uf} that the closed topological string
partition function can be expressed in terms of genus zero Gopakumar-Vafa
invariants as 
\begin{equation}
{Z}_{\mathrm{top}}^X(q,Q)=M(q)^{\chi \left( X\right) /2}\,
\prod\limits_{\beta \in H_2(X,\mathbb{Z})^+}~\prod_m\,
M(q^m\,Q^\beta,q)^{-N_{m}^{\beta }}  \label{Ztop}
\end{equation}%
where $\chi(X)$ is the topological Euler characteristic of $X$, the integer $%
N_m^\beta$ counts the number of BPS states of M2-branes in curve class $%
\beta=(\beta_1,\dots,\beta_s)\in H_2(X,\mathbb{Z})$ and with intrinsic $%
SU(2) $ spin $m$, and $Q^\beta=Q_1^{\beta_1}\cdots Q_s^{\beta_s}$. This
expression describes the generating functions $Z_{\mathrm{BPS}}^X$, counting
BPS bound states of a single D6 brane wrapping $X$ with D2 and D0
branes, in the large radius chamber. On the other hand, in the chamber
corresponding to the noncommutative point in the K\"ahler moduli
space, the BPS
partition function is given by~\cite%
{Szendroi:2007nu,Jafferis:2008uf,Chuang:2008aw}
\begin{eqnarray}
Z_{\mathrm{BPS}}^X={Z}_{\mathrm{top}}^X(q,Q)\, {Z}_{\mathrm{top}%
}^X(q,Q^{-1}) \ .  \label{ZBPSX}
\end{eqnarray}
The appropriate restriction of (\ref{ZBPSX}), which depends on the
value of the $B$-field as described in e.g.~\cite{Jafferis:2008uf},
describes BPS states in various chambers of the K\"ahler moduli space which are separated by walls
of marginal stability where BPS states decay or form.

The symmetric function expansion in this case involves the other
generalization of the Schur polynomials, the Jack polynomials $\mathfrak{j}%
_{\lambda }^{\left( \alpha \right) }(x)$ \cite{Stanleyp,macdonald} with $%
\mathfrak{j}_{\lambda }^{\left( \alpha =1\right) }(x)=\mathfrak{s}_{\lambda
}(x)$. They satisfy a Cauchy identity%
\begin{equation}
\sum_{\lambda }\, c_{\lambda }\left( \alpha \right) \, \mathfrak{j}_{\lambda
}^{\left( \alpha \right) }(x)\, \mathfrak{j}_{\lambda }^{\left( \alpha
\right) }(y)=\prod\limits_{i,j\geq 1}\, \left( 1-x_{i}\, y_{j}\right)
^{-1/\alpha } \ ,  \label{CauchyJack}
\end{equation}%
where $c_{\lambda }\left( \alpha \right) $ are rational functions of the
parameter $\alpha$ which have been calculated in \cite{Stanleyp}. Due to the
power $1/\alpha $ in (\ref{CauchyJack}), they are well suited to describe
the topological string partition function as written in (\ref{Ztop}). The
generic expression (\ref{Ztop}) can thus be written in terms of the Jack
polynomials as%
\begin{equation}
\frac{{Z}^X_{\mathrm{top}}(q,Q)}{M(q)^{\chi \left( X\right) /2}}%
=\prod\limits_{\beta\in H_2(X,\mathbb{Z})^+}~\prod_m~ \sum_{\lambda }\,
c_{\lambda }\big(-1/N_m^\beta\big) \, q^{m\,|\lambda|}\, Q^{\left\vert
\lambda \right\vert\,\beta }\, \mathfrak{j}_{\lambda }^{( -1/N_{m}^{\beta })
}(q^{i-1/2})^2 \ ,
\end{equation}%
and the powers of the MacMahon function can be expressed as 
\begin{equation}
M(q)^{\chi \left( X\right) /2}=\sum_{\lambda }\, c_{\lambda }\big(2/\chi(X) %
\big)\, \mathfrak{j}_{\lambda }^{(2/\chi(X))}(q^{i-1/2})^2 \ .
\end{equation}

For the backgrounds considered in this paper, the (unreduced) noncommutative
partition functions are related to those of $X$ by the wall-crossing factor $%
W^X=\widetilde{Z}_{\mathrm{DT}}^X(q,Q^{-1})$, which describes the crossing
of an infinite number of walls in going from the noncommutative point to the
large volume point~\cite{Szendroi:2007nu,Jafferis:2008uf}. We can index the
walls crossed by $i\in\mathbb{N}$ and factorize the wall-crossing factor as $%
W^X=\prod_{i\geq1}\,W_i^X$, with $W_i^X$ the jump of the BPS state partition
function across the $i$-th wall. This yields the partition function (\ref%
{ZBPSX}) in the $\ell$-th chamber 
\begin{eqnarray}
Z_{\mathrm{BPS}}^X(\ell)=\mathcal{W}_\ell^X\,Z_{\mathrm{DT}}^X(q,Q) \ ,
\label{ZBPSXn}
\end{eqnarray}
where $\mathcal{W}_\ell^X=\prod_{i\geq \ell}\,W_i^X$.

For the resolved conifold $X={\mathbb{X}}$, this prescription yields $W_i^{%
\mathbb{X}}=(1-(-q)^i)^i$ and 
\begin{eqnarray}
\mathcal{W}_\ell ^{\mathbb{X}}=\big((-q)^\ell \,Q^{-1}\,;\,-q\big)_\infty^\ell \, 
\widetilde{Z}_{\mathrm{DT}}^{\,{\mathbb{X}}}\big(q\,,\,(-q)^\ell \,Q^{-1}\big)
\end{eqnarray}
where $(a;q)_\infty\equiv \prod_{j\in\mathbb{N}}\,(1-a\,q^j)$. The expansion
of (\ref{ZBPSXn}) in Hall-Littlewood functions is given by 
\begin{eqnarray}
Z_{\mathrm{BPS}}^{\mathbb{X}}(\ell ) &=& \big((-q)^\ell \,Q^{-1}\,;\,-q\big)%
_\infty^\ell \, \sum_{\lambda }\, b_{\lambda }\left(Q\right) \, \mathfrak{p}%
_{\lambda }\big(\left( -q\right) ^{i-{1}/{2}};Q \big)^{2}  \notag \\
&& \hspace{4cm} \times \, \sum_{\mu }\, b_{\mu }\big((-q)^\ell \,
Q^{-1}\big)\, \mathfrak{p}_{\mu }\big(\left( -q\right) ^{i-{1}/{2}%
};(-q)^\ell \, Q^{-1} \big)^{2} \ .
\end{eqnarray}
This function interpolates between $Z_{\mathrm{BPS}}^{\mathbb{X}}(1)=Z_{%
\mathrm{DT}}^{\mathbb{A}}$ (after reparametrization) and $Z_{\mathrm{BPS}}^{%
\mathbb{X}}(\infty)=Z_{\mathrm{DT}}^{\mathbb{X}}$, at the noncommutative and
large volume limit points, through values of $\ell $ corresponding to chambers
in the K\"ahler cone of $\mathbb{X}$. For the resolution of the $A_1$
quotient singularity $X={\mathbb{Y}}$, using duality one has $\mathcal{W}_\ell ^{%
\mathbb{Y}}=\big(\mathcal{W}_\ell ^{\mathbb{X}}\big)^{-1}$. The expansion of the
BPS state partition function (\ref{ZBPSXn}) in Schur functions is then given
by 
\begin{eqnarray}
Z_{\mathrm{BPS}}^{\mathbb{Y}}(\ell )&=& \big((-q)^\ell
\,Q^{-1}\,;\,-q\big)_\infty^\ell 
\notag \\
&& \times\, \sum_{\lambda,\mu,\nu }\, Q^{\left\vert \mu \right\vert
-|\nu|}\,(-q)^{\ell \,|\nu|}\, \mathfrak{s}_{\lambda }\big((-q)^{i-1/2}\big)%
^{2}\, \mathfrak{s}_{\mu }\big((-q)^{i-1/2}\big)^{2}\, \mathfrak{s}_{\nu }%
\big((-q)^{i-1/2}\big)^{2} \ .  \label{ZBPSYn}
\end{eqnarray}

\subsection{Generalizations}

These constructions generalize in a natural way to all local toric
Calabi-Yau threefolds without compact divisors. Let us briefly illustrate
this through a representative set of examples, generalizing the conifolds
and $\mathbb{Z}_2$-orbifolds considered above. Let $G$ be a finite subgroup
of $SU(2)$ acting on $\mathbb{C}^3$ via the natural embedding $SU(2)\subset
SU(3)$. The crepant resolution of the orbifold $\mathbb{C}^3/G$ given by the 
$G$-Hilbert scheme of $\mathbb{C}^3$ is ${\mathbb{Y}}_G= {\mathbb{S}}%
_G\times \mathbb{C}$, where ${\mathbb{S}}_G$ is the minimal ADE resolution
of the double point singularity $\mathbb{C}^2/G$. By the McKay
correspondence~\cite{Szabo:2009vw}, the non-trivial irreducible $G$%
-representations correspond to simple roots of an associated ADE root system 
$\Delta$, the collection of which gives a basis for $H_2({\mathbb{Y}}_G,%
\mathbb{Z})$. Let $\Delta^+\subset\Delta$ be the set of positive roots, and $%
n$ the number of irreducible representations of $G$. Then the
Donaldson-Thomas partition function generalizes (\ref{OrbY}) to~\cite{AGYJ} 
\begin{eqnarray}
Z_{\mathrm{DT}}^{{\mathbb{Y}}_G}(q,Q)=M(-q)^n\, \prod_{\beta\in\Delta^+}\,
M(Q^\beta,-q) \ .  \label{ZDTYG}
\end{eqnarray}
The noncommutative Donaldson-Thomas invariants arising from the McKay quiver
associated to the affine ADE Dynkin diagram are the same as the orbifold
Donaldson-Thomas invariants of $\mathbb{C}^3/G$~\cite{JoyceSong,AGYJ}. The
corresponding partition function $Z_{\mathrm{DT}}^{\mathbb{C}^3/G}(q,Q)$ is
given by the same formula (\ref{ZDTYG}) but with the product now ranging
over the full root lattice $\Delta$~\cite{AGYJ}. Hence the corresponding
reduced partition functions can be expanded in Schur functions as 
\begin{eqnarray}
\widetilde{Z}_{\mathrm{DT}}^{\,{\mathbb{Y}}_G}\left(
q,Q\right)=\prod_{\beta\in\Delta^+}~ \sum_{\lambda }\, Q^{\left\vert \lambda
\right\vert\,\beta }\,\mathfrak{s}_{\lambda }\big((-q)^{i-1/2}\big)^{2} \ ,
\label{ZDTYGSchur}
\end{eqnarray}
and the identical formula for $\widetilde{Z}_{\mathrm{DT}}^{\,\mathbb{C}%
^3/G}\left( q,Q\right)$ involving a product over all roots. The
corresponding BPS state partition functions in the various chambers can be
worked out as before.

For the special case $G=\mathbb{Z}_n$, the products run over roots of the $%
A_{n-1}$ Lie algebra and the partition function (\ref{ZDTYG}) specializes to~%
\cite{Young,Szendroi:2007nu} 
\begin{eqnarray}
Z_{\mathrm{DT}}^{{\mathbb{Y}}_{\mathbb{Z}_n}}(q,Q)=M(-q)^n\,\prod_{1\leq
i\leq j<n}\, M(Q_{[i,j]},-q) \ ,  \label{ZDTYZn}
\end{eqnarray}
where $Q_{[i,j]}\equiv Q_i\,Q_{i+1}\cdots Q_j$. This is (after
reparametrization) the generating function for $n$-coloured plane
partitions. It is ``dual'' to the Donaldson-Thomas partition function for
the generalized conifold geometry ${\mathbb{X}}_n$~\cite%
{Iqbal:2004ne,Nagao,Aganagic:2009kf} which is given by 
\begin{eqnarray}
Z_{\mathrm{DT}}^{{\mathbb{X}}_{n}}(q,Q)=M(-q)^n\,\prod_{1\leq i\leq j<n}\,
M(Q_{[i,j]},-q)^{N_{ij}} \ ,  \label{ZDTXn}
\end{eqnarray}
where $N_{ij}=-(-1)^{n_{ij}}$ with $n_{ij}$ the number of internal edges
between vertices $i$ and $j$ in the toric web diagram for ${\mathbb{X}}_n$.
The corresponding reduced function is expanded in Schur functions as 
\begin{eqnarray}
\widetilde{Z}_{\mathrm{DT}}^{\,{\mathbb{X}}_n}(q,Q)&=& \prod_{\overset{%
\scriptstyle 1\leq i\leq j<n}{\scriptstyle N_{ij}=+1}}~\sum_{\lambda }\,
(Q_{[i,j]})^{\left\vert \lambda \right\vert }\,\mathfrak{s}_{\lambda }\big(%
(-q)^{k-1/2}\big)^{2}  \notag \\
&& \times\, \prod_{\overset{\scriptstyle 1\leq i\leq j<n}{\scriptstyle %
N_{ij}=-1}} ~\sum_{\mu }\, (-Q_{[i,j]})^{\left\vert \mu \right\vert }\, 
\mathfrak{s}_{\mu }\big((-q)^{k-1/2}\big)\, \mathfrak{s}_{\mu^{\prime }}\big(%
(-q)^{k-1/2}\big) \ .  \label{ZDTXnSchur}
\end{eqnarray}

These expressions should all follow from the representions of the
Donaldson-Thomas partition functions as correlators of vertex
operators given in~\cite{Sulkowski:2009rw}, as it is known that such
correlation functions can be represented in terms of Schur functions. 
These formulas also formally apply to the $\mathbb{C}^3/\mathbb{Z}_2\times%
\mathbb{Z}_2$ orbifold and its symmetric resolution, the closed topological
vertex geometry. The noncommutative chamber was originally considered
in~\cite{Young}, while several infinite families of chambers and their
wall-crossing
behaviours are identified in~\cite{Sulkowski:2009rw}. They can also be extended to some non-toric
geometries without compact divisors, such as the formal toric Calabi-Yau
threefolds considered in~\cite{Hollowood:2003cv,Caporaso:2006kk} (see also~%
\cite{Aganagic:2009kf}). The resulting expressions are straightforward but
somewhat tedious to write down, and one must be somewhat more careful with
the convergence of the various infinite products involved. We omit the
details.

\section{Unitary matrix models for Donaldson-Thomas theory\label{DTMM}}

\subsection{Chern-Simons theory and Toeplitz determinants\label{CSToeplitz}}

The interest of Toeplitz determinants in physics begins with Onsager's study
of the two-dimensional Ising model, as he succeeded in showing that the
diagonal spin correlation function is given by an $N\times N$ Toeplitz
determinant~\cite{Onsager,Onsageralso}. Another classic example is the
one-plaquette model of two-dimensional Yang-Mills theory with gauge group $%
U(N)$, whose partition function is given by \cite{Bars:1979xb}%
\begin{equation}
Z_{N}(\lambda)=\det_{1\leq i,j\leq N} \,\big[ I_{i-j}(2\lambda)\big] \ ,
\label{IT}
\end{equation}%
where $I_{n}(z)$ is the modified Bessel function order $n$. The same problem
was shortly afterwards studied using unitary one-matrix models \cite{GW},
leading to the well-known Gross-Witten matrix model 
\begin{equation}
Z_{N}(\lambda)=\prod\limits_{j=1}^{N}~\int_{0}^{2\pi }\, \frac{\mathrm{d}%
\theta_j}{2\pi}~ \exp \left( \lambda\, \big({\,\mathrm{e}}\,^{{\,\mathrm{i}\,%
}\theta_j}+{\,\mathrm{e}}\,^{-{\,\mathrm{i}\,}\theta_j}\big)%
\right)~\prod\limits_{k<l}\,\big\vert {\,\mathrm{e}}\,^{{\,\mathrm{i}\,}%
\theta _{k}}-{\,\mathrm{e}}\,^{{\,\mathrm{i}\,}\theta _{l}}\big\vert ^{2} {\
\ .}  \label{GW}
\end{equation}%
In recent work~\cite{Szabo:2010qv} we showed that the unitary matrix model
for $U(N)$ Chern-Simons gauge theory on the three-sphere $S^3$ is, in a
certain sense, a $q$-deformation of the Gross-Witten model. Here we shall
show that its partition function can also be represented as a Toeplitz
determinant.

The reason for the existence of two equivalent representations is a direct
relationship between Toeplitz determinants and unitary random matrix models,
found long before the introduction of matrix models. It is given by the
Heine-Szeg\"{o} identity \cite{Szego}%
\begin{equation}
\prod\limits_{j=1}^{N}~\int_{0}^{2\pi }\, \frac{\mathrm{d}\theta_j}{2\pi}~ f%
\big( {\,\mathrm{e}}\,^{{\,\mathrm{i}\,}\theta _{j}}\big)~
\prod\limits_{k<l}\,\big\vert {\,\mathrm{e}}\,^{{\,\mathrm{i}\,}\theta _{k}}-%
{\,\mathrm{e}}\,^{{\,\mathrm{i}\,}\theta _{l}}\big\vert ^{2}=D_{N}\left(f%
\right) \ ,  \label{Heine}
\end{equation}%
where $D_{N}(f)$ is the $N\times N$ Toeplitz determinant with symbol $f$,%
\begin{equation}
D_{N}(f)=\det_{1\leq j,k\leq N}\, \left[\widehat{f}(j-k)\right] \ ,
\label{T}
\end{equation}%
with $\widehat{f}(r)=\frac{1}{2\pi }\, \int_{0}^{2\pi }\,
f(\e^{\ii\theta})\, {\,\mathrm{e}}%
\,^{{\,\mathrm{i}\,} r\, \theta }~\mathrm{d}\theta$, $r\in\mathbb{Z}$ the
Fourier coefficients of the symbol function. The symbol function $f(z)$ is
the weight function of the corresponding matrix model; we assume throughout
that it lives in $C^\infty(S^1)$. In the case of the Gross-Witten model, one
uses the identity%
\begin{equation}
\exp \Big(\lambda\, \big(z+z^{-1}\big)\Big) =\sum_{n=-\infty }^{\infty }\,
I_{n}(2\lambda)\, z^n
\end{equation}%
to establish the equivalence of the two expressions $\left( \ref{IT}\right) $
and $\left( \ref{GW}\right) .$

Let us now compute the $U(N) $ Chern-Simons free energy on $S^{3}$ in terms
of a Toeplitz determinant$.$ The unitary matrix model is given by the
partition function \cite{Okuda:2004mb} 
\begin{equation}
Z_{\mathrm{CS}}^{U(N)}\left( S^{3}\right) \equiv
\prod\limits_{j=1}^{N}~\int_{0}^{2\pi }\, \frac{\mathrm{d}\theta_j}{2\pi}%
~\Theta({\,\mathrm{e}}\,^{{\,\mathrm{i}\,}\theta
_{j}}|q)~\prod\limits_{k<l}\,\big\vert {\,\mathrm{e}}\,^{{\,\mathrm{i}\,}%
\theta _{k}}-{\,\mathrm{e}}\,^{{\,\mathrm{i}\,}\theta _{l}}\big\vert ^{2}
\end{equation}%
with 
\begin{equation}
\Theta(z|q)=\sum_{j=-\infty}^\infty\, q^{j^{2}/2}\, z^j \ .  \label{theta}
\end{equation}%
We have seen that the entries of the Toeplitz determinant $\left( \ref{T}%
\right) $ are given by the Fourier coefficients of the symbol, which is the
weight function of the unitary matrix model. In this case, the symbol is the
theta-function $\Theta(z|q)$ and the entries of the matrix are automatically
at our disposal. Then the determinant is%
\begin{equation}
D_{N}(\Theta)=%
\begin{vmatrix}
a_{0} & a_{1} & \cdots & a_{N-1} \\ 
a_{-1} & a_{0} & \cdots & a_{N-2} \\ 
\vdots & \vdots &  & \vdots \\ 
a_{-N+1} & a_{-N+2} & \cdots & a_{0}%
\end{vmatrix}%
\qquad \text{ with } \quad a_{j}=q^{{j^{2}}/{2}} \ .  \label{D}
\end{equation}%
Hence the required determinant is $\det_{i,j} \, \big[q^{{\left( i-j\right)
^{2}}/{2}}\big],$ which is known to give rise to the Chern-Simons partition
function \cite{dHT}%
\begin{equation}
Z_{\mathrm{CS}}^{U(N)}\big(S^3\big)= \det_{1\leq i,j\leq N}\, \big[q^{{%
\left( i-j\right)^{2}}/{2}}\big] =\prod\limits_{k=1}^{N-1}\, \big( 1-q^{k}%
\big) ^{N-k} \ .  \label{ZCSdet}
\end{equation}%
Notice that the non-trivial part of the determinant, $\det_{i,j}\,
[q^{i\,j}],$ is just a Vandermonde determinant.

A more general Toeplitz determinant, with symbol 
\begin{equation}
\sum_{j=0}^{\infty }\,q^{{{j^{2}}/{2}}}\, z^{j}  \label{trunc}
\end{equation}%
was already computed in \cite{LS}. It was proven there, by induction, that
the determinant $\left( \ref{D}\right) $ with coefficient $a_{m}$ instead of 
$a_{0}$ is given by%
\begin{equation}
D_{m,N}=\big( q^{m\,(m-1)/2}\big) ^{N}\, \prod\limits_{k=1}^{N-1}\, \big( %
1-q^{k}\big) ^{N-k} \ .
\end{equation}%
They consider the case $m>N$ because their symbol is $\left( \ref{trunc}%
\right) $, but if the symbol is the full theta-function $\left( \ref{theta}%
\right) $ we can take $m=0$ and recover $D_{0,N}=Z_{\mathrm{CS}}^{U(N)}(S^3)$
given in (\ref{ZCSdet}).

Of course, knowledge of the Toeplitz determinant result directly implies,
due to (\ref{Heine}), that there is a unitary matrix model representation
with a theta-function as weight function, a model found by other means in 
\cite{Okuda:2004mb}. In \cite{Tierz} it was shown that the inverse of the
theta-function also leads to a viable Chern-Simons matrix model. We shall
see that there is an equivalence between these results and the ensuing
representation of Donaldson-Thomas partition functions in terms of unitary
matrix models.

\subsection{Donaldson-Thomas theory on $\mathbb{C}^3$\label{C3matrix}}

As in the case of Chern-Simons gauge theory on $S^3$, we can write the
partition function of Donaldson-Thomas theory on $\mathbb{C}^3$ as a
Toeplitz determinant. For this, we use Gessel's formula for the product of
Schur polynomials in terms of a Toeplitz determinant, which in terms of the
Schur measure reads~\cite{Gessel}%
\begin{equation}
\mathcal{P}_N(x,y)\equiv \sum_{\lambda \, : \, \lambda _{1}\leq N}\, {%
\mathfrak{s}}_{\lambda }\left( x\right) \, {\mathfrak{s}}_{\lambda }\left(
y\right) =D_{N}(A_{i-j}) \ ,  \label{Toeplitz}
\end{equation}%
where 
\begin{equation}
A_{i}=A_{i}(x,y)=\sum_{l=0}^{\infty }\, \mathfrak{h}_{l+i}(x)\, \mathfrak{h}%
_{l}(y) \ ,
\end{equation}%
and $\mathfrak{h}_{r}(x)=\sum_i\,x_i^r$ is the $r$-th complete symmetric
function. Then the Donaldson-Thomas partition function is given by%
\begin{equation}
Z_{\mathrm{DT}}^{%
\mathbb{C}
^{3}}\left( q\right) =\lim_{N\rightarrow \infty }\, \mathcal{P}_N\big(%
x_{i}=q^{i-1/2}\,,\,y_i =q^{i-1/2}\big) \ .
\end{equation}

It is convenient to consider the symbol of the Toeplitz determinant \cite%
{Gessel,TW} 
\begin{equation}
f_{\mathbb{C}^3} \left( z\right) =\sum_{i=-\infty }^{\infty }\,
A_{i}(x,y)\,z^{i}=\prod\limits_{j\geq1}\,\left( 1-y_{j}\, z^{-1}\right)
^{-1}\,\left( 1-x_{j}\, z\right) ^{-1} \ .
\end{equation}%
Taking into account the Heine-Szeg\"{o} identity (\ref{Heine}), the
Donaldson-Thomas partition function can then be written as the $N\to\infty$
limit of an $N\times N$ matrix model 
\begin{eqnarray}
Z_{\mathrm{DT}}^{%
\mathbb{C}
^{3}}\left( q\right) &=&\lim_{N\rightarrow \infty
}\,\prod\limits_{n=1}^{N}~\int_0^{2\pi}\, \frac{\mathrm{d}\theta_n}{2\pi}%
~\prod\limits_{j=1}^{\infty }\, \left( 1-q^{j-1/2}~{\,\mathrm{e}}\,^{-{\,%
\mathrm{i}\,}\theta _{n}}\right) ^{-1}\, \left( 1-q^{j-1/2}~{\,\mathrm{e}}%
\,^{{\,\mathrm{i}\,}\theta _{n}}\right) ^{-1}  \notag \\
&& \hspace{4cm} \times~ \prod\limits_{k<l}\, \big\vert {\,\mathrm{e}}\,^{{\,%
\mathrm{i}\,}\theta _{k}}-{\,\mathrm{e}}\,^{{\,\mathrm{i}\,}\theta _{l}}%
\big\vert ^{2} \ .
\end{eqnarray}%
Using the Jacobi triple product formula for the theta-function%
\begin{equation}
\Theta(z|q) =\prod\limits_{j=1}^{\infty }\,\left(1-q^{j}\right)\, \left(
1+q^{j-1/2}\, z^{-1}\right)\, \left( 1+q^{j-1/2}\, z\right) \ ,
\label{triple}
\end{equation}%
we can write 
\begin{equation}
Z_{\mathrm{DT}}^{%
\mathbb{C}
^{3}}\left( q\right) =A_{\infty }\,\prod\limits_{n=1}^{\infty
}~\int_{0}^{2\pi }\,\frac{\mathrm{d}\theta _{n}}{2\pi }~ \frac{1}{\Theta( -{%
\,\mathrm{e}}\,^{{\,\mathrm{i}\,}\theta _{n}}|q) }~ \prod\limits_{k<l}\, %
\big\vert {\,\mathrm{e}}\,^{{\,\mathrm{i}\,}\theta _{k}}-{\,\mathrm{e}}\,^{{%
\,\mathrm{i}\,}\theta _{l}}\big\vert ^{2} \ ,  \label{mat1}
\end{equation}%
where $A_{\infty }=\lim_{N\rightarrow \infty }\,\left(q\right) _{\infty
}^{N} $. Thus the Donaldson-Thomas partition function is a unitary $N=\infty 
$ one-matrix model with weight function $w \left( \theta\right) =\Theta( -{\,%
\mathrm{e}}\,^{{\,\mathrm{i}\,}\theta}|q)^{-1} .$

Recall that the matrix model for $U(N)$ Chern-Simons gauge theory on $S^{3}$
is a unitary matrix model with weight function $w^{\prime }\left( \theta
\right) =\Theta( {\,\mathrm{e}}\,^{{\,\mathrm{i}\,}\theta }|q) $ and $N$
eigenvalues \cite{Okuda:2004mb,Szabo:2010qv}. We relate the two models and
the corresponding partition functions in a more precise way below. Before
doing that, we can also demonstrate the relationship using an equivalent
matrix model representation of the Donaldson-Thomas partition function,
based again on the Cauchy identity but now written as 
\begin{equation}
\sum_{\lambda }\, {\mathfrak{s}}_{\lambda ^{\prime }}\left( x\right)\, {%
\mathfrak{s}}_{\lambda ^{\prime }}\left( y\right) =\prod\limits_{i,j\geq
1}\, \frac{1}{1-x_{i}\, y_{j}}
\end{equation}%
where $\lambda ^{\prime }$ are the conjugate (transposed) partitions. The
symbol in this case is \cite{TW} 
\begin{equation}
f_{\mathbb{C}^3}^{\prime }\left( z\right) =\prod\limits_{j\geq1}\, \left(
1+y_{j}\, z^{-1}\right) \,\left( 1+x_{j}\,z\right) \ ,  \label{symbol2}
\end{equation}%
and the corresponding matrix model representation is%
\begin{eqnarray}
Z_{\mathrm{DT}}^{%
\mathbb{C}
^{3}}\left( q\right) &=&\lim_{N\rightarrow \infty }\,
\prod\limits_{n=1}^{N}~\int_{0}^{2\pi }~\frac{\mathrm{d}\theta_n}{2\pi}%
~\prod\limits_{j=1}^{\infty }\,\left( 1+q^{j-1/2}~ {\,\mathrm{e}}\,^{-{\,%
\mathrm{i}\,}\theta _{n}}\right)\, \left( 1+q^{j-1/2}~{\,\mathrm{e}}\,^{{\,%
\mathrm{i}\,}\theta _{n}}\right)  \notag \\
&& \hspace{4cm} \times~ \prod\limits_{k<l}\, \big\vert {\,\mathrm{e}}\,^{{\,%
\mathrm{i}\,}\theta _{k}}-{\,\mathrm{e}}\,^{{\,\mathrm{i}\,}\theta _{l}}%
\big\vert ^{2} \ .
\end{eqnarray}%
Using (\ref{triple}) again, the expression in terms of theta-functions is
then 
\begin{equation}
Z_{\mathrm{DT}}^{%
\mathbb{C}
^{3}}\left( q\right) =C_{\infty }\,\prod\limits_{n=1}^{\infty
}~\int_{0}^{2\pi }\, \frac{\mathrm{d}\theta _{n}}{2\pi }~ \Theta( {\,\mathrm{%
e}}\,^{{\,\mathrm{i}\,}\theta _{n}}|q) ~ \prod\limits_{k<l}\, \big\vert {\,%
\mathrm{e}}\,^{{\,\mathrm{i}\,}\theta _{k}}-{\,\mathrm{e}}\,^{{\,\mathrm{i}\,%
}\theta _{l}}\big\vert ^{2} \ ,  \label{Matrixmodel}
\end{equation}%
with $C_{\infty }=\lim_{N\rightarrow \infty }\, \left(q\right) _{\infty
}^{-N}.$ While we use a different definition for the theta function, this result 
is identical to the one previously found in \cite{Ooguri:2010yk}. For the case of the conifold and 
orbifold Donaldson-Thomas theories we will find some differences between our matrix 
models and the ones in \cite{Ooguri:2010yk}. This may be due to the non-uniqueness of the matrix 
model description.

\subsection{Chern-Simons/Donaldson-Thomas correspondence}

The two matrix model representations, (\ref{mat1}) and (\ref{Matrixmodel}),
parallels the situation in Chern-Simons theory, since the weight function $w
\left( \theta \right) =\Theta( -{\,\mathrm{e}}\,^{{\,\mathrm{i}\,}%
\theta}|q)^{-1} $ can also be used for a Chern-Simons matrix model. The
Chern-Simons unitary matrix model representation thus not only matches (\ref%
{Matrixmodel}) but also (\ref{mat1}). This result appeared in \cite[eq.~(4.8)%
]{Tierz}, where the non-uniqueness of the matrix model representation was
pointed out, based on the undetermined moment problem for the
Stieltjes-Wigert weight function. It follows from the explicit expression,
given by Askey, for a weight function that has the same moments as the
log-normal distribution, which is given by the inverse of a theta-function~%
\cite{Askey}%
\begin{equation}
w_\gamma \left(z\right) =c_{\gamma }\, \frac{z^{\gamma -1}}{\left(
-z;q\right) _{\infty }\, \left( -q/z;q\right) _{\infty }} \ .
\end{equation}%
If we set $z=\sqrt{q}~{\,\mathrm{e}}\,^{{\,\mathrm{i}\,}\theta }$ and choose 
$\gamma =0$, then by the triple product formula (\ref{triple}) we directly
obtain the Chern-Simons --- or finite $N$ --- matrix model version of (\ref%
{mat1}).\footnote{%
In \cite{Tierz}, the particular case $\gamma =-3/2$ was considered, which
gives one of the most common normalizations of the Stieltjes-Wigert
polynomials.}

Thus in both cases, we see that the right-hand side is essentially the
matrix model for $U(N)$ Chern-Simons gauge theory on $S^{3}$ when $N=\infty
. $ Indeed, considering the corresponding generating functions, we have%
\begin{eqnarray}
Z_{\mathrm{CS}}^{U(N)}(S^{3}) =\prod\limits_{j=1}^{N-1}\, \frac{(1-q^{j})^{N}%
}{(1-q^{j})^{j}} \qquad \mbox{and} \qquad Z_{\mathrm{DT}}^{%
\mathbb{C}
^{3}}\left( q\right) =\prod\limits_{j=1}^{\infty }\, \frac{1}{\left(
1-q^{j}\right) ^{j}} \ .  \label{Zs}
\end{eqnarray}%
Then $\left( \ref{Matrixmodel}\right) $ implies $Z_{\mathrm{DT}}^{%
\mathbb{C}
^{3}}\left( q\right) =C_{\infty }\, \lim_{N\rightarrow \infty }\, Z_{\mathrm{%
CS}}^{U(N)}(S^{3}),$ which also follows from $\left( \ref{Zs}\right) $ and 
\begin{equation}
\left(q\right) _{\infty }^{-N}=\prod\limits_{j=1}^{\infty }\, \left(
1-q^{j}\right) ^{-N} \ .
\end{equation}%
From the very definitions of their partition functions, the Donaldson-Thomas
free energy $F_{\mathrm{DT}}^{\mathbb{C}^3}=\log Z_{\mathrm{DT}}^{\mathbb{C}%
^3}$ and the Chern-Simons free energy $F_{\mathrm{CS}}^{U(N)}=\log Z_{%
\mathrm{CS}}^{U(N)}$ at $N=\infty $ thus satisfy the simple relationship%
\begin{equation}
F_{\mathrm{DT}}^{%
\mathbb{C}
^{3}}\left( q\right) =F_{\mathrm{CS}}^{U(\infty )}(S^{3})-\lim_{N\rightarrow
\infty }\, N\, \sum_{j=1}^{\infty }\, \log \left( 1-q^{j}\right) \ .
\label{eqF}
\end{equation}%
Hence, via a precise infinite renormalization, the Donaldson-Thomas free
energy is exactly the free energy of $U(N)$ Chern-Simons gauge theory in the
limit $N\to \infty $.

The factor $\lim_{N\rightarrow \infty }\,\left(q\right) _{\infty }^{-N}$
that links the two partition functions has a natural interpretation. As we
have seen, the $q$-Pochammer symbol%
\begin{equation}
\left(q\right) _{n}=\prod\limits_{j=1}^{n}\, \left( 1-q^{j}\right)
\end{equation}%
is the building block of the Chern-Simons partition function. In the limit $%
n\rightarrow \infty $, it also appears as the partition function of $U(N)$
topologically twisted Vafa-Witten $\mathcal{N}=4$ gauge theory \cite%
{Vafa:1994tf} on $%
\mathbb{C}
^{2}$, which is given by \cite{Szabo:2009vw} 
\begin{equation}
Z_{U(N)}^{%
\mathbb{C}
^{2}}\left( q\right) =(q)_{\infty }^{-N} \ .  \label{ZVW}
\end{equation}%
We can thus write 
\begin{equation}
Z_{\mathrm{DT}}^{%
\mathbb{C}
^{3}}\left( q\right) =\lim_{N\rightarrow \infty }\, Z_{U(N)}^{%
\mathbb{C}
^{2}}\left( q\right) \, Z_{\mathrm{CS}}^{U(N)}(S^{3}) \ .  \label{ZDTVWCS}
\end{equation}%
It would be interesting to better understand this relationship from the
point of view of the six-dimensional $U(1)$ topological gauge theory
underlying the Donaldson-Thomas invariants~\cite{INOV,CSS}. If we regard $%
\mathbb{C}^3$ as the trivial line bundle over $\mathbb{C}^2$, then this
result suggests that the fibre degrees of freedom in the six-dimensional
gauge theory can be integrated out, leaving its natural gauge theory
counterpart on $\mathbb{C}^2$ and on the boundary $S^3$, at infinite rank.
This sort of reduction of the gauge theory partition function is
demonstrated in the case of local toric surfaces in~\cite{BT}.

However, in (\ref{ZVW}) the (complexified) gauge coupling parameter is $%
q=\exp (2\pi {\,\mathrm{i}\,}\tau )$ with $\tau =\frac{4\pi {\,\mathrm{i}\,}%
}{g_{\mathrm{YM}}^{2}}$ (for vanishing $\theta$-angle). If we use the
relationship with the topological string coupling constant $g_{s}=g_{\mathrm{%
YM}}^{2}/2$, then the quantum parameter is $q=\exp (-4\pi ^{2}/g_{s})$,
while the quantum parameter in Chern-Simons theory is $q=\exp \left(
-g_{s}\right) .$ Thus in (\ref{ZDTVWCS}) the Chern-Simons partition function
should be understood in its dual form after performing a Gauss summation.
This is the correct setting for merging the four-dimensional Vafa-Witten
theory with its corresponding boundary Chern-Simons gauge theory~\cite%
{GSST,BT}. Indeed, the equality (\ref{ZDTVWCS}) can be regarded as the large 
$N$ limit of the partition function for $q$-deformed Yang-Mills theory on
the two-sphere $S^2$~\cite{GSST}, which describes the relationship between
the two gauge theories. We will discuss the relation between
Donaldson-Thomas theory and $q$-deformed gauge theories in the next section.

\subsection{Conifold Donaldson-Thomas theory}

The case of the conifold partition functions can be similarly studied. Using 
$\mathfrak{s}_{\lambda }(Q\, x)=Q^{|\lambda|}\, \mathfrak{s}_{\lambda }(x)$,
the symbol associated to the slightly more general expansion $\left( \ref%
{dual}\right)$ is given by 
\begin{equation}
f_{\,{\mathbb{X}}} \left( z\right) =\prod\limits_{j\geq1}\, \left(
1-Q\,y_{j} \, z^{-1}\right) ^{-1}\, \left( 1+x_{j}\, z\right) \ .
\end{equation}%
Hence the Donaldson-Thomas partition function of the resolved conifold
(without degree zero contributions) has the matrix model representation 
\begin{equation}
\widetilde{Z}_{\mathrm{DT}}^{\,{\mathbb{X}}}\left( q,Q\right)
=\lim_{N\rightarrow \infty }\, \prod\limits_{n=1}^{N}~ \int_{0}^{2\pi }\, 
\frac{\mathrm{d}\theta _{n}}{2\pi }~ \prod\limits_{j=1}^{\infty }\,\frac{%
1+q^{j-1/2}~ {\,\mathrm{e}}\,^{{\,\mathrm{i}\,}\theta _{n}}}{1-Q\, q^{j-1/2}~%
{\,\mathrm{e}}\,^{-{\,\mathrm{i}\,}\theta _{n}}}~ \prod\limits_{k<l}\big
\vert {\,\mathrm{e}}\,^{{\,\mathrm{i}\,}\theta _{k}}-{\,\mathrm{e}}\,^{{\,%
\mathrm{i}\,}\theta _{l}}\big\vert ^{2} \ .
\end{equation}

The noncommutative conifold partition function (\ref{ZA}) can be written as
the product of two Toeplitz determinants $D(f)=\det T(f)$, where $T(f)=\big[%
\,\widehat{f}(j-k)\big]_{j,k\geq1}$ is the corresponding $\ell^2(\mathbb{N})$
Toeplitz operator. However, the product of two Toeplitz operators is not of
Toeplitz form~\cite{Treview}. To deal with this problem, one considers all
Toeplitz matrices $T_N(f)\equiv \Pi_N\,T(f)\, \Pi_N$ for $N\in\mathbb{N}$
together, where $\Pi_N$ is the orthogonal projection onto the Fourier modes $%
1,\dots,N$. Then a sequence of products of Toeplitz matrices $\left\{
T_{N}\left( f\right)\, T_{N}\left(g\right) \right\}_{N\in\mathbb{N}} $ is
asymptotically equivalent to the sequence of Toeplitz matrices $\left\{
T_{N}\left( f\, g\right) \right\}_{N\in\mathbb{N}} $ \cite{Treview}. The
requisite condition is that the Fourier coefficients of the symbol $t_k=%
\widehat{f}(k) $ (the entries of the Toeplitz matrix) are absolutely
summable 
\begin{equation}
\sum_{k=-\infty }^{\infty }\, \vert t_k\vert <\infty \ .  \label{absum}
\end{equation}%
A sequence of Toeplitz matrices $T_{N}=[t_{j-k}]_{1\leq j,k\leq N}$ for
which the $t_{k}$ are absolutely summable is said to be in the Wiener class.
Likewise, a function $f\left(z \right) $ defined on the circle $S^1$ is said
to be in the Wiener class if it has a Fourier series expansion with
absolutely summable Fourier coefficients. If the symbols are of Wiener
class, then one has the convergence result~\cite{Treview} 
\begin{equation}
\lim_{N\rightarrow \infty }\,\big\Vert T_{N}\left(f\right)\, T_{N}\left(
g\right) -T_{N}\left( f\, g\right) \big\Vert = 0
\end{equation}%
where $\|-\|$ denotes the operator norm on finite-dimensional matrices.

In our case, the symbols are given by theta-functions which have Fourier
coefficients of the type $t_{k}=q^{{{k^{2}}/{2}}}$ with $0<q<1$. Absolute
summability is consequently satisfied and hence the symbols are of Wiener
class. The limit $N\rightarrow \infty $ is precisely the one that we are
studying, and hence we can write 
\begin{eqnarray}
\widetilde{Z}_{\mathrm{DT}}^{\,\mathbb{A}}\left( q,Q\right) &=&
\lim_{N\rightarrow \infty }\,\prod\limits_{n=1}^{N}~ \int_{0}^{2\pi }\, 
\frac{\mathrm{d}\theta _{n}}{2\pi }~\prod\limits_{j=1}^{\infty }\,\frac{%
\left( 1+q^{j-1/2}~{\,\mathrm{e}}\,^{{\,\mathrm{i}\,}\theta _{n}}\right) ^{2}%
}{\left( 1-{Q}\, q^{j-1/2}~{\,\mathrm{e}}\,^{-{\,\mathrm{i}\,}\theta
_{n}}\right)\, \left( 1-{Q}^{-1}\,q^{j-1/2}~{\,\mathrm{e}}\,^{-{\,\mathrm{i}%
\,}\theta _{n}}\right) }  \notag \\
&& \hspace{4cm} \times~ \prod\limits_{k<l}\, \big\vert {\,\mathrm{e}}\,^{{\,%
\mathrm{i}\,}\theta _{k}}-{\,\mathrm{e}}\,^{{\,\mathrm{i}\,}\theta _{l}}%
\big\vert^{2} \ ,  \label{ZA-Mat}
\end{eqnarray}%
with the limit understood in the sense of norm convergence.

\subsection{Orbifold Donaldson-Thomas theory}

The orbifold partition functions are treated analogously. The symbol
associated to the Schur function expansion $\left( \ref{inv}\right) $ is%
\begin{equation}
f_{\,{\mathbb{Y}}} \left( z\right) =\prod\limits_{j\geq1}\, \left( 1-Q\,
y_{j}\, z^{-1}\right) ^{-1}\, \left( 1-x_{j}\, z\right) ^{-1} \ .
\end{equation}%
This leads to the matrix model representation 
\begin{eqnarray}
{\widetilde{Z}_{\mathrm{DT}}^{\,{\mathbb{Y}}}\left( q,Q\right) }
&=&\lim_{N\rightarrow \infty }\,\prod\limits_{n=1}^{N}~ \int_{0}^{2\pi }\,%
\frac{\mathrm{d}\theta _{n}}{2\pi }~ \prod\limits_{j=1}^{\infty }\, \frac{1}{%
\left( 1-Q\, q^{j-1/2}~{\,\mathrm{e}}\,^{-{\,\mathrm{i}\,}\theta
_{n}}\right)\, \left( 1-q^{j-1/2}~{\,\mathrm{e}}\,^{{\,\mathrm{i}\,}\theta
_{n}}\right) }  \notag \\
&& \hspace{4cm} \times~ \prod\limits_{k<l}\,\big\vert {\,\mathrm{e}}\,^{{\,%
\mathrm{i}\,}\theta _{k}}-{\,\mathrm{e}}\,^{{\,\mathrm{i}\,}\theta _{l}}%
\big\vert^{2} \ .
\end{eqnarray}%
Likewise, by using the same arguments which led to (\ref{ZA-Mat}), the
noncommutative orbifold partition function (\ref{invorb}) admits the matrix
model representation 
\begin{eqnarray}
\widetilde{Z}_{\mathrm{DT}}^{\,\mathbb{C}^3/\mathbb{Z}_2}\left( q,Q\right)
&=& \lim_{N\rightarrow \infty }\,\prod\limits_{n=1}^{N}~ \int_{0}^{2\pi }\,%
\frac{\mathrm{d}\theta _{n}}{2\pi }~ \prod\limits_{j=1}^{\infty }\, \frac{%
\left( 1-q^{j-1/2}~{\,\mathrm{e}}\,^{{\,\mathrm{i}\,}\theta _{n}}\right)^{-2}%
}{\left( 1-Q\, q^{j-1/2}~{\,\mathrm{e}}\,^{-{\,\mathrm{i}\,}\theta
_{n}}\right)\, \left( 1-Q^{-1}\, q^{j-1/2}~{\,\mathrm{e}}\,^{-{\,\mathrm{i}\,%
}\theta _{n}}\right) }  \notag \\
&& \hspace{4cm} \times~ \prod\limits_{k<l}\,\big\vert {\,\mathrm{e}}\,^{{\,%
\mathrm{i}\,}\theta _{k}}-{\,\mathrm{e}}\,^{{\,\mathrm{i}\,}\theta _{l}}%
\big\vert^{2} \ .
\end{eqnarray}

As pointed out in~\cite{Treview}, the arguments which led to (\ref{ZA-Mat})
easily extend to the more general case of products of $m>2$ Toeplitz
matrices. In particular, one has the convergence results 
\begin{eqnarray}
\lim_{N\rightarrow \infty }\,\big\Vert T_{N}\left(f_1\right)\cdots
T_{N}\left( f_m\right) -T_{N}\left( f_1\cdots f_m\right) \big\Vert = 0 \ .
\end{eqnarray}
This enables us to write down matrix model representations for all
generalizations considered in the previous section. For example, the BPS
state partition function (\ref{ZBPSYn}) can be expressed as 
\begin{eqnarray}
Z_{\mathrm{BPS}}^{\mathbb{Y}}(\ell )&=& \big((-q)^\ell \,Q^{-1}\,;\,-q\big)%
_\infty^\ell \, C_\infty\, \lim_{N\to\infty}\, \prod\limits_{m=1}^{N}~
\int_{0}^{2\pi }\,\frac{\mathrm{d}\theta _{m}}{2\pi }~ \Theta({\,\mathrm{e}}%
\,^{{\,\mathrm{i}\,}\theta_m}|q)  \notag \\
&& \times~ \prod\limits_{j=1}^{\infty }\, \frac{\left( 1-q^{j-1/2}~{\,%
\mathrm{e}}\,^{{\,\mathrm{i}\,}\theta _{m}}\right)^{-2}}{\left( 1-Q\,
q^{j-1/2}~{\,\mathrm{e}}\,^{-{\,\mathrm{i}\,}\theta _{m}}\right)\, \left(
1-Q^{-1}\,(-q)^\ell \, q^{j-1/2}~{\,\mathrm{e}}\,^{-{\,\mathrm{i}\,}\theta
_{m}}\right) } ~ \prod\limits_{k<l}\,\big\vert {\,\mathrm{e}}\,^{{\,\mathrm{i%
}\,}\theta _{k}}-{\,\mathrm{e}}\,^{{\,\mathrm{i}\,}\theta _{l}}\big\vert^{2}
\ .
\end{eqnarray}
Similarly, the generalized orbifold partition function (\ref{ZDTYGSchur}) is 
\begin{eqnarray}
&& \\
\widetilde{Z}_{\mathrm{DT}}^{\,{\mathbb{Y}}_G}\left( q,Q\right)=
\lim_{N\to\infty}\, \prod\limits_{n=1}^{N}~ \int_{0}^{2\pi }\,\frac{\mathrm{d%
}\theta _{n}}{2\pi }~ \prod\limits_{j=1}^{\infty }~ \prod_{\beta\in\Delta^+}
\, \frac{\left( 1-q^{j-1/2}~{\,\mathrm{e}}\,^{{\,\mathrm{i}\,}\theta
_{n}}\right)^{-1}}{1-Q^\beta\, q^{j-1/2}~{\,\mathrm{e}}\,^{-{\,\mathrm{i}\,}%
\theta _{n}}} ~ \prod\limits_{k<l}\,\big\vert {\,\mathrm{e}}\,^{{\,\mathrm{i}%
\,}\theta _{k}}-{\,\mathrm{e}}\,^{{\,\mathrm{i}\,}\theta _{l}}\big\vert^{2}
\ ,  \notag
\end{eqnarray}
while the generalized conifold partition function (\ref{ZDTXn}) can be
expressed as 
\begin{eqnarray}
\widetilde{Z}_{\mathrm{DT}}^{\,{\mathbb{X}}_n}(q,Q)&=&\lim_{N\to\infty}\,
\prod\limits_{m=1}^{N}~ \int_{0}^{2\pi }\,\frac{\mathrm{d}\theta _{m}}{2\pi }%
~ \prod\limits_{p=1}^{\infty }~ \prod_{\overset{\scriptstyle 1\leq i\leq j<n}%
{\scriptstyle N_{ij}=+1}}\, \frac{\left( 1-q^{p-1/2}~{\,\mathrm{e}}\,^{{\,%
\mathrm{i}\,}\theta _{m}}\right)^{-1}}{1-Q_{[i,j]}\, q^{p-1/2}~{\,\mathrm{e}}%
\,^{-{\,\mathrm{i}\,}\theta _{m}}}  \notag \\
&& \times ~ \prod_{\overset{\scriptstyle 1\leq i\leq j<n}{\scriptstyle %
N_{ij}=-1}}\, \frac{1+q^{p-1/2}~ {\,\mathrm{e}}\,^{{\,\mathrm{i}\,}\theta
_{m}}}{1-Q_{[i,j]}\, q^{p-1/2}~{\,\mathrm{e}}\,^{-{\,\mathrm{i}\,}\theta
_{m}}}~\prod\limits_{k<l}\,\big\vert {\,\mathrm{e}}\,^{{\,\mathrm{i}\,}%
\theta _{k}}-{\,\mathrm{e}}\,^{{\,\mathrm{i}\,}\theta _{l}}\big\vert^{2} \ ,
\end{eqnarray}
and so on. Notice that, in contrast to the $\mathbb{C}^3$ case, the matrix models in 
this and the previous Section are very similar but show some differences 
with the matrix models obtained in \cite{Ooguri:2010yk}. We expect to address 
this issue in future work.

The Toeplitz determinant can also be expressed as a Fredholm determinant 
\cite{GC,BO}, which suggests a deep underlying integrability structure, as
we discuss in detail in the next section. This applies directly to the
Chern-Simons partition function, which is given by a finite $N\times N$
Toeplitz determinant, for which the result of~\cite{BO} immediately applies.
This result does not strictly apply in the limit $N\to\infty$ which yields
the Donaldson-Thomas partition function. In this limit, the Fredholm
determinant representation converges to $1$ and the infinite Toeplitz
determinant is given by the normalization constant of the Schur measure~\cite%
{BO}. This agrees with our computation, wherein the matrix model
representation of Donaldson-Thomas theory follows directly from the Toeplitz
determinant representation of the Cauchy identity (\ref{Toeplitz}) and the
Heine-Szeg\"o identity (\ref{Heine}). Proper specialization of the symmetric
functions then yields the Donaldson-Thomas partition functions. For the
first part of this procedure, one can alternatively use the results of \cite%
{BR}, where the Cauchy identity is expressed directly as a generic matrix
model average. Other instances where the partition function can be written
as a Fredholm determinant are the partition function of two-dimensional
quantum gravity~\cite{Moore:1990mg} and the grand canonical partition
function of $c=1$ string theory with vortex excitations~\cite{Kazakov:2000pm}%
.

\section{Integrability structure\label{DTint}}

\subsection{Toda and Toeplitz lattice hierarchies}

The expansions in Schur functions are also useful to establish a
relationship between Donaldson-Thomas theory and the theory of integrable
hierarchies, a connection that is generally expected whenever a matrix model
formulation is available~\cite{Gerasimov:1990is,Krichever:1992qe}. Using the
results of~\cite{AvM}, it is straightforward to identify the
Donaldson-Thomas partition functions as particular instances of a
tau-function of the 2-Toda lattice hierarchy \cite{2Toda}. More precisely,
they are given by tau-functions of the Toeplitz lattice hierarchy~\cite%
{AvM,AvM2}, which is a reduction of the 2-Toda lattice hierarchy.\footnote{%
As pointed out in \cite{AvM}, the Toeplitz lattice hierarchy is better known
as the Ablowitz-Ladik hierarchy~\cite{AbLa} which arises in discretizations
of the non-linear Schr\"odinger equation, but we follow the terminology of 
\cite{AvM,AvM2}. The other reduction of the 2-Toda lattice hierarchy leads
to the standard Toda lattice hierarchy \cite{AvM}. The Toeplitz reduction
(on $S^1$) and the Toda reduction (on $\mathbb{R}$) are essentially
equivalent~\cite{Cafasso}.} The tau-functions of the 2-Toda lattice
hierarchy $\tau _{n}(t,s)$, $n\in 
\mathbb{Z}
$ depend on two sets of time variables $t,s\in \mathbb{C}^{\infty }$ and are
defined by the Hirota bilinear equations. They can also be written as the
determinant of a semi-infinite moment matrix%
\begin{equation}
\tau _{n}(t,s)=\det m_{n}(t,s) \ .
\end{equation}%
In the case of the reduction to the Toeplitz lattice hierarchy, this moment
matrix is a Toeplitz matrix (whereas in the reduction to the standard Toda
lattice hierarchy it is a Hankel matrix), and the components of the vector $%
\tau (t,s)=(\tau _{0}(t,s)=1,\tau _{1}(t,s),\dots)$ of tau-functions of the
Toeplitz lattice hierarchy satisfy~\cite{AvM}%
\begin{equation}
\tau _{n}(t,s)=\sum_{\lambda\, :\,\lambda _{1} \leq n}\, \mathfrak{s}%
_{\lambda }(t)\, \mathfrak{s}_{\lambda }(-s) \ .  \label{tau-adler}
\end{equation}

Hence if the times are taken to be $t_{i}=q^{i-1/2}$ and $s_{j}=-q^{j-1/2},$
then%
\begin{equation}
\tau _{\infty }(q^{i-1/2},-q^{j-1/2})=Z_{\mathrm{DT}}^{%
\mathbb{C}
^{3}}\left( q\right) \ .
\end{equation}%
On the other hand, if the two sets of time variables are taken to be $t_{i}=%
\sqrt{Q}\, q^{i-1/2}$ and $s_{j}=-\sqrt{Q}\, q^{j-1/2},$ then%
\begin{equation}
\tau _{\infty }(\sqrt{Q}\, q^{i-1/2},-\sqrt{Q}\, q^{j-1/2})= \widetilde{Z}_{%
\mathrm{DT}}^{\,\mathbb{Y}}(q,Q)=\widetilde{Z}_{\mathrm{DT}}^{\,\mathbb{X}%
}(q,Q)^{-1}.
\end{equation}%
Thus while one begins within the formalism of the 2-Toda lattice hierarchy,
the Donaldson-Thomas partition functions are a particular case of a
reduction of the 2-Toda lattice hierarchy. This reduction is equivalent to
the one-dimensional Toda lattice hierarchy, a result that has also been
shown in the context of the melting crystal picture \cite{Nakatsu:2007dk}.
While our partition functions involve two sets of Schur polynomials, the two
sets of times are taken to be equal. That this reduces the system to a
one-dimensional Toda hierarchy is shown in \cite{Nakatsu:2007dk}. Using the
free fermionic representation of a deformation of the expansion (\ref{inv}),
they show that it is a 2-Toda tau-function that satisfies%
\begin{equation}
\tau\left( t,s\right) =\tau\left( t-s,0\right) \ ,
\end{equation}%
and thus reduces to a one-dimensional Toda lattice hierarchy.

More generally, the tau-function of the two-component KP hierarchy is also a
function of two sets of time variables and admits the double Schur function
expansion~\cite{Sato}%
\begin{equation}
\tau_{\mathrm{KP}}\left(t,s\right) =\sum_{\lambda ,\mu }\, c_{\lambda \mu
}\, \mathfrak{s}_{\lambda }(t)\, \mathfrak{s}_{\mu }(-s) \ ,
\end{equation}%
where $c_{\lambda \mu }$ are Pl\"{u}cker coordinates of a point on a
two-component analog of the Sato grassmannnian~\cite{Sato} and have
well-known determinant expressions. (The Schur expansion for tau-functions
of the ordinary KP hierarchy consists of only one set of Schur functions.)
The case of the 2-Toda lattice hierarchy follows from this one, as it only
has an additional discrete index $n$ as in (\ref{tau-adler}). More
precisely, following~\cite{HO} these tau-functions can be written as images
of the Pl\"ucker map corresponding to projection along a basis element of
the charge $N$ sector of a fermionic Fock space as 
\begin{equation}
\tau _{N,g}^{(2)}(t,s)=\sum_{\lambda ,\mu }\, B_{N,g}\left( \lambda ,\mu
\right)\, \mathfrak{s}_{\lambda }(t)\, \mathfrak{s}_{\lambda}(-s) \ .
\label{2-Toda}
\end{equation}%
Here $B_{N,g}\left( \lambda ,\mu \right) $ are Pl\"{u}cker coordinates of
the image of an element $g\in Gr_{\mathcal{H}_{+}^{N}}(\mathcal{H})$ in the
grassmannian of subspaces of the Hilbert space $\mathcal{H}=L^2(S^1)$ whose
orthogonal projections onto the subspace $\mathcal{H}_+\subset\mathcal{H}$,
consisting of functions that admit holomorphic extensions to the interior of 
$S^1\subset\mathbb{C}$, have Fredholm index $N$. Then the noncommutative
Donaldson-Thomas partition function $\widetilde{Z}_{\mathrm{DT}}^{\,\mathbb{A%
}}\left( q,Q\right) $ is a particular case of (\ref{2-Toda}) with unit
Pl\"ucker coordinates.

The particular case of (\ref{2-Toda}) for the identity element $g=e$ is \cite%
{HO} 
\begin{equation}
\tau _{N,e}^{(2)}(t,s)=\sum_{\lambda }\, \mathfrak{s}_{\lambda }(t)\, 
\mathfrak{s}_{\lambda}(-s)
\end{equation}%
which, with the specification of the time variables given above, describes
the partition functions $Z_{\mathrm{DT}}^{%
\mathbb{C}
^{3}}\left( q\right) $ and $\widetilde{Z}_{\mathrm{DT}}^{\,\mathbb{Y}}(q,Q)$%
, and reduces to a one-dimensional Toda lattice tau-function. With the
matrix model expression for $\widetilde{Z}_{\mathrm{DT}}^{\,\mathbb{A}%
}\left( q,Q\right) $, one can interpret it as the product of two
one-dimensional Toda tau-functions. The more general partition functions
described earlier are likewise given as products of 1-Toda tau-functions.

\subsection{Isomonodromic tau-functions}

Following \cite{ITW}, a somewhat more conjectural relationship with
tau-functions and integrable systems can also be established, using the fact
that the Donaldson-Thomas and Chern-Simons partition functions are Toeplitz
determinants. The result of~\cite{ITW} shows that some Toeplitz determinants
can be interpreted as certain isomonodronic tau-functions, which coincide
with those of Jimbo, Miwa and Ueno \cite{iso}. They focus on a particular
case of the Gessel identity (\ref{Toeplitz}) where one set of variables $%
x=(x_1,\dots,x_n)$ is generic and left unspecified, while the other set has
the specification $y=(1,\dots,1)$. In this case the associated symbol is%
\begin{equation}
f_{\mathrm{JMU}}\left( z\right) ={\,\mathrm{e}}\,^{1/z}\,
\prod\limits_{j=1}^{n}\, \left( 1+x_{j}\, z\right) \ .
\end{equation}
However, as pointed out in \cite{ITW}, their derivation of integrable
partial differential equations can be applied to any Toeplitz determinant
for which the logarithmic derivative of the associated symbol $f(z)$ is a
rational function of $z$.

The Chern-Simons partition functions for gauge group $U(N)$ have
finite-dimensional Toeplitz determinant representation and symbol given by a
theta-function (\ref{triple}) after the proper specification, i.e. $%
n\rightarrow \infty$. This symbol violates the rational behaviour required
of its logarithmic derivative, unless we truncate the products at some large
value $M<n$. The extension to infinite support of the variables $x=(x_i)$ is
left as an open problem in \cite{ITW}. In this ``finite-dimensional''
approximation, the calculation of the Toeplitz determinant in Section~\ref%
{CSToeplitz} proceeds based on the Fourier coefficients up to $a_{M-1}$, and
corresponds to a truncation of the symbol to index $M$. Then the truncated
Chern-Simons partition functions can be identified with the Jimbo-Miwa-Ueno
tau-functions corresponding to the (generalized) Schlesinger isomonodromy
deformation equations of the $2\times 2$ linear matrix ordinary differential
equations which have $M$ simple poles in $\mathbb{C}$ and one irregular
singular point at infinity of Poincar\'{e} index one.

The rigorous proof that the Chern-Simons partition function is an
isomonodromic tau-function is beyond the scope of the present paper.%
\footnote{%
This proof presumably follows the arguments indicated in~\cite{Eynard:2010dh}
that the essential characteristics are independent of the cutoff $M$ on the
length of the representation associated to $\lambda$.} It is then an
interesting open problem to show that the Donaldson-Thomas partition
functions are tau-functions of a similar system of non-linear partial
differential equations, which have both infinitely many poles in the complex
plane and are given by infinite-dimensional Toeplitz determinants. The
isomonodronic tau-functions are known~\cite{Palmer} to be intimately related
to the KP tau-functions discussed above and introduced in~\cite{Sato}, hence
it should be possible to further establish a stronger relationship between
these two families of integrable hierarchies and Donaldson-Thomas
theory. The expressions of generic Donaldson-Thomas partition
functions as correlators involving exponentials of fermionic bilinears
given in~\cite{Sulkowski:2009rw} shows that they automatically provide
tau-functions of integrable hierarchies.

\subsection{Nekrasov functions}

The expansion of Donaldson-Thomas partition functions in terms of the $q$%
-Plancherel measure is also useful for establishing a relationship with $%
\mathcal{N}=2$ supersymmetric gauge theory in four dimensions with Casimir
operators, since Nekrasov's formulas \cite{Nek} can be also written in terms
of the Plancherel measure \cite{NekOuk,LMN}. This sort of relationship has
already been noticed, within the crystal melting picture in~\cite%
{Maeda:2004iq,Nakatsu:2007dk} and also in~\cite{Tai:2007vc} within a
slightly different context. In particular, for the noncommutative $U(1)$
gauge theory\footnote{%
In general, one considers an abelian $\mathcal{N}=2$ gauge theory with
instantons. The noncommutative gauge theory is one possibility, whereby the
field theory is embedded in the $\Omega$-backgound with toric deformation
parameters $(\epsilon_1,\epsilon_2)$. Other possibilities are the gauge
theories of fractional D3-branes at an ADE singularity or of the D5/NS5
brane system wrapping a $\mathbb{P}^{1}$ in $K3$ \cite{LMN}.} the instanton
partition function $Z_{\mathrm{inst}}(a,t,\epsilon _{1},\epsilon _{2})$ can
be expressed as a sum over partitions \cite{NekOuk,LMN}. In the Calabi-Yau
case when $\epsilon _{1}=-\epsilon _{2}=\hslash $, one has 
\begin{equation}
Z_{\mathrm{inst}}(a,{t},\hslash)=\sum_{\lambda }\, \frac{(\dim\lambda)^{2}}{%
(-\hslash ^{2})^{\left\vert \lambda \right\vert }}\, \exp \Big(- \frac{1}{%
\hslash ^{2}}\, \sum_{k\geq1}\, t_{k}\, \frac{\mathrm{ch}_{k+1}\left(
a,\lambda \right) }{k+1}\, \Big) \ ,  \label{Nek}
\end{equation}%
where $t=(t_k)$ are coupling constants, $\mathrm{ch}_{k+1}\left( a,\lambda
\right) $ are the Chern characters of the partition $\lambda$, and $a$ is
the vacuum expectation value of the vector multiplets. Thus it coincides
with the principal specification\footnote{%
Principal means that it utilizes Schur functions (infinite number of
variables), as in Donaldson-Thomas theory.} of the Schur functions%
\begin{equation}
\dim\lambda=\lim_{q\rightarrow 1}\, \lim_{N\rightarrow \infty }\, \mathfrak{s%
}_{\lambda }(1,q,\dots,q^{N-1}) \ .
\end{equation}

As in \cite{NekOuk,LMN}, let us consider the case $t_{1}\neq 0$ and $%
t_{2}=t_{3}=\dots=0$, i.e. all higher Casimir operators are turned off.
Using the Chern character $\mathrm{ch}_{2}\left( a,\lambda \right)
=a^{2}+2\hslash ^{2}\left\vert \lambda \right\vert $, in that case we have%
\begin{equation}
Z_{\mathrm{inst}}(a,{t}_1,\hslash)=\sum_{\lambda }\, \frac{\mathfrak{s}%
_{\lambda }(1,1,\dots)^2}{(-\hslash ^{2})^{\left\vert \lambda \right\vert }}%
\, \exp \Big(- \frac{t_{1}}{\hslash ^{2}}\, a^{2}-t_{1}\left\vert \lambda
\right\vert \, \Big) \ .  \label{simpleNek}
\end{equation}%
Consider now the $\mathbb{C}^3/\mathbb{Z}_2$ orbifold Donaldson-Thomas
partition function (\ref{inv}), and take its $q\to1$ limit using 
\begin{equation}
\lim_{q\rightarrow 1}\, \mathfrak{s}_{\lambda
}(q^{i-1/2})^{2}=g_{s}^{-2\left\vert \lambda \right\vert }\, \mathfrak{s}%
_{\lambda }(1,1,\dots)^{2}
\end{equation}%
where $q={\,\mathrm{e}}\,^{-g_s}$. Then the partition function $\widetilde{Z}%
_{\mathrm{DT}}^{\,\mathbb{Y}}(-q,Q)$ in the limit $q\rightarrow 1$ is the
non-equivariant limit of Nekrasov's partition function with $Q={\,\mathrm{e}}%
\,^{-t_{1}}$ and $g_{s}={\,\mathrm{i}\,}\hslash \rightarrow 0$, 
\begin{equation}
\lim_{q\rightarrow 1}\, \widetilde{Z}_{\mathrm{DT}}^{\,\mathbb{Y}}(-q=-{\,%
\mathrm{e}}\,^{-{\,\mathrm{i}\,}\hslash},Q={\,\mathrm{e}}\,^{-t_1})={\,%
\mathrm{e}}\,^{t_{1}\,a^{2}/2\hslash^{2}}\, Z_{\mathrm{inst}}(a,{t},\hslash)
\ .
\end{equation}

The equivariant case which is directly related with Donaldson-Thomas theory
is that of the five-dimensional $\mathcal{N}$ = 1 supersymmetric $U(1)$
gauge theory (or K-theory version of the original gauge theory), because its
instanton expansion involves the $q$-Plancherel measure instead of the
Plancherel measure above \cite{Maeda:2004iq,Nakatsu:2007dk}.\footnote{%
The string coupling $g_s$ here is then interpreted as the circumference of
the compactified fifth dimension.} Its partition function coincides with $%
\widetilde{Z}_{\mathrm{DT}}^{\,\mathbb{Y}}(q,Q) $ and $\widetilde{Z}_{%
\mathrm{DT}}^{\,\mathbb{X}}(q,Q)^{-1}$ given in (\ref{inv})~\cite%
{Maeda:2004iq}. As we have seen above and as is also shown explicitly in~%
\cite{Nakatsu:2007dk}, this partition function is a tau-function of the
one-dimensional Toda lattice hierarchy and not of the two-dimensional Toda
lattice hierarchy.

\subsection{Hurwitz theory}

Another gauge theory partition function that can be immediately written in
terms of the Plancherel measure is the heat kernel expansion of
two-dimensional Yang-Mills theory~\cite{Migdal75,Rusakov}. On the sphere $%
S^{2}\cong\mathbb{P}^1$, from this expansion or from the matrix model
representation~\cite{Gross:1994mr} the partition function can be written in
terms of Schur polynomials as%
\begin{equation}
{Z}_{\mathrm{YM}}^{U(N)}\left(\mathbb{P}^1\right) =\sum_{\lambda }\, 
\mathfrak{s}_{\lambda }(1,\dots,1)^{2}\, q^{C_{2}(\lambda)} \ ,  \label{Z-2d}
\end{equation}%
where $\mathfrak{s}_{\lambda }(1,\dots,1)=\dim\lambda$ is the dimension of
the irreducible representation of the $U(N)$ gauge group corresponding to $%
\lambda=(\lambda_1,\dots,\lambda_N) $, and $C_{2}(\lambda )$ is the
quadratic Casimir invariant of the representation. Hence in the large $N$
limit, whereby the gauge group is $U\left( \infty \right),$ it coincides
with the Nekrasov partition function (\ref{Nek}) with only the first two
Casimir operators turned on, an observation already made by~\cite{NekOuk}
(see also~\cite{Tai:2007vc}).

The large $N$ limit picks out the chiral sector of the two-dimensional gauge
theory which receives contributions from only ``small'' representations $%
\lambda$ to the partition function (\ref{Z-2d}), weighted by the linear
Casimir operator. Then the $U(\infty )$ chiral partition function is related
to the Nekrasov function (\ref{simpleNek}), and one has 
\begin{equation}
{Z}_{\mathrm{YM}}^{U(\infty )}(\mathbb{P}^1)^{+}= \lim_{q\rightarrow 1}\, 
\widetilde{Z}_{\mathrm{DT}}^{\,\mathbb{Y}}\left( -q,Q\right) .
\end{equation}%
Of course, the $q$-deformation of this theory is more intimately related to
Donaldson-Thomas theory, as then no limit is required. We shall discuss this
case in more detail below. We first point out that the expansion of the
partition function in terms of Schur polynomials allows us to apply results
of Okounkov \cite{OkounToda} and relate the partition functions with
tau-functions of the Toda lattice hierarchy, while it further enables the
study of their role as generating functions for ramified coverings of $%
\mathbb{P}^1$ \cite{coverings}.

Let us begin by briefly summarizing the Toda lattice results of \cite%
{OkounToda}. Let $H_{d}(C_{1},\dots,C_{s})$ denote the weighted number of
(possibly disconnected) $d$-fold coverings of $\mathbb{P}^{1}$, ramified
over $s$ fixed points of $\mathbb{P}^{1}$ with monodromies in fixed
conjugacy classes $C_{1},\dots,C_{s}$ of the symmetric group $S_d$.\footnote{%
The genus $g$ of the covering surface is determined in terms of the
branching structure and the degree $d$ by the Riemann-Hurwitz formula.} The
generating function for degree $d$ coverings of $\mathbb{P}^{1}$ whose
ramification over $0,\infty \in \mathbb{P}^{1}$ can be fixed arbitrarily
while the remaining $b=s-2$ ramifications are simple is given by 
\begin{equation}
Z_{\mathrm{Hur}}^{(2)}\left( x,y, q,Q\right) =\sum_{d,b=0}^\infty \, Q^{d}\, 
\frac{(-g_s) ^{b}}{b!}~ \sum_{|\mu|=|\nu|=d }\, x_{\mu }\, y_{\nu
}~H_{d}(C_{\mu },C_{\nu },\underset{\text{$b$ times}}{\underbrace{%
C_{(2)},\dots,C_{(2)}}}) \ ,  \label{tauokoun}
\end{equation}%
where $C_\mu$ is the conjugacy class corresponding to the partition $\mu$ of 
$d$, $C_{(2)}$ is the conjugacy class of a transposition, and $%
x_\mu=\prod_i\, x_{\mu_i}$ for a set of variables $x=(x_1,x_2,\dots)$. The
generating function for connected coverings is the corresponding free energy%
\begin{equation}
F_{\mathrm{Hur}}^{(2)}\left( x,y, v ,q\right) =\log Z_{\mathrm{Hur}%
}^{(2)}\left( x,y, v ,q\right) =\sum_{d,b=0}^\infty \, Q^{d}\, \frac{(-g_s)
^{b}}{b!}~ \sum_{|\mu|=|\nu|=d }\, x_{\mu }\, y_{\nu }~H^\bullet_{d,b}\left(
\mu ,\nu \right) \ ,
\end{equation}%
where $H^\bullet_{d,b}\left( \mu ,\nu \right) $ denote the double Hurwitz
numbers introduced by Okounkov which are the weighted numbers of connected
degree $d$ coverings of $\mathbb{P}^{1}$ with monodromy around $0,\infty \in 
\mathbb{P}^{1}$ given by $%
\mu
$ and $\nu $, respectively, and $b$ are the numbers of additional simple
ramifications.

The difference between simple Hurwitz numbers and double Hurwitz numbers is
as follows. Let $%
\mu
$ be a partition of an integer $d$. Then the simple Hurwitz number $h_{\mu
}^{g}$ is the number of connected genus $g$ branched covers of $\mathbb{P}%
^{1}$ with monodromy given by $%
\mu
$ over a fixed point, usually identified with $\infty $, and an appropriate
number of fixed simple branch points. In the more general setting of double
Hurwitz numbers, there is a useful explicit expression for the partition
function (\ref{tauokoun}) given by 
\begin{equation}
Z_{\mathrm{Hur}}^{(2)}\left( x,y, v ,q\right) =\sum_{\lambda }\,
Q^{\left\vert \lambda \right\vert }~q^{ \kappa(\lambda )/2}\, \mathfrak{s}%
_{\lambda }(x)\, \mathfrak{s}_{\lambda }(y) \ ,  \label{Oktau}
\end{equation}%
which is proven in \cite{OkounToda} to be a tau-function of the Toda lattice
hierarchy. Here $\kappa(\lambda )$ is one of the shifted symmetric
polynomials that has the explicit expression 
\begin{equation}
\kappa(\lambda )=\sum_{i}\, \lambda_i\,\left( \lambda _{i}-2i+1\right) \ ,
\end{equation}%
which is essentially the Casimir eigenvalue $C_{2}\left( \lambda \right) $ 
\cite{OkounOlshan}. It is then clear that (\ref{Oktau}) can be specified to
describe (\ref{Z-2d}) simply by choosing $x=y=Q^{-1/2}\, (1,\dots,1).$ This
implies that the Yang-Mills partition function ${Z}^{U(N)}_{\mathrm{YM}%
}\left( \mathbb{P}^1\right) $ is a tau-function of the Toda lattice
hierarchy.

Once again, the $q$-deformed case is the one more directly connected to
Donaldson-Thomas theory, and the formula (\ref{Oktau}) applies to that case
as well via a suitable specification of $x=(x_{1},x_{2},\dots)$ and $%
y=(y_{1},y_2,\dots)$. The heat kernel expansion for the partition function
of $q$-deformed two-dimensional Yang-Mills theory on the sphere can be
written in terms of Schur polynomials as~\cite{Aganagic:2004js} 
\begin{equation}
{Z}_{q-\mathrm{YM}}^{U(N)}\left( \mathbb{P}^1\right) =\sum_{\lambda }\, 
\mathfrak{s}_{\lambda }(1,q,\dots,q^{N-1})^{2}\, q^{p\, C_{2}(\lambda )} \ ,
\label{Zq}
\end{equation}%
where $\mathfrak{s}_{\lambda }(1,q,\dots,q^{N-1}) = \dim_q\lambda$ is the
quantum dimension of the $U(N)$ representation associated to $\lambda$. As
above, it can be interpreted as a tau-function of the Toda lattice hierarchy
and it is also a generating function for double Hurwitz numbers. As before,
due to the argument of the quadratic versus linear behaviour of the
exponential term in (\ref{Zq}), it is the partition function of the $%
U(\infty )$ chiral sector of $q$-deformed Yang-Mills theory that can be
related to the orbifold partition function $\widetilde{Z}_{\mathrm{DT}}^{\,%
\mathbb{Y}}(q^p,Q)$ (or to the instanton partition function of the K-theory
version of the gauge theory above).

\subsection{Equivariant Gromov-Witten theory}

According to the work of Gross and Taylor \cite%
{Gross:1992tu,Gross:1993hu,Gross:1993yt}, two-dimensional Yang-Mills theory
has a large $N$ string expansion based on branched coverings of $\mathbb{P}%
^1 $. The Hurwitz theory involved is the one that enumerates simple Hurwitz
numbers \cite{Gross:1993hu,Gross:1993yt,Caporaso:2006gk}. In the formalism
of double Hurwitz numbers, it corresponds to taking $x_{2}=x_{3}=\dots=0$
above. On the other hand, in the application of \cite{OkounToda} to
two-dimensional Yang-Mills theory and its $q$-deformation on $\mathbb{P}^1$
the two sets of variables $x=y$ are taken to be equal, in order to describe
the $\mathfrak{s}_{\lambda }^{2}$ terms of the partition functions (as
occurs throughout this paper). Hence only a particular case of (\ref{Oktau})
is required, and this seems to indicate that it is more natural to consider
two branch points at $0$ and $\infty $, instead of just one, with the same
monodromies. This is true for both the ordinary and the $q$-deformed case,
changing only the branching data which depends on the parameter $q$ in the
latter case.

The $q$-deformed gauge theory conjecturally describes a non-perturbative
completion of the A-model topological string theory on the local toric
Calabi-Yau threefold given by the total space of the rank two holomorphic
bundle ${\mathbb{X}}_p=\mathcal{O}_{\mathbb{P}^1}(p-2)\oplus \mathcal{O}_{%
\mathbb{P}^1}(-p)$ over $\mathbb{P}^{1}$ (generalizing the resolved conifold 
${\mathbb{X}}_1={\mathbb{X}}$ and the $\mathbb{C}^3/\mathbb{Z}_2$ orbifold
resolution ${\mathbb{X}}_2={\mathbb{Y}}$). The closed perturbative
topological string partition function on ${\mathbb{X}}_{p}$ is given by \cite%
{Aganagic:2004js,Caporaso:2006gk}%
\begin{equation}
Z_{\mathrm{top}}^{{\mathbb{X}}_{p}}(q,Q)=\sum_{\lambda }\, (-1)^{p\,
|\lambda|}\, Q^{\left\vert \lambda \right\vert}\,q^{\left( p-2\right)\,
\kappa(\lambda)/2}\, W_{\lambda }(q)^{2} \ ,
\end{equation}%
where 
\begin{equation}
W_{\lambda }(q)=q^{-\kappa(\lambda)/4}\, \prod\limits_{u\in \lambda }\, 
\frac{1}{\left[ h(u)\right] }
\end{equation}%
is a specialization of the topological vertex~\cite{Aganagic:2003db}. In
terms of the principal $q$-specialization of the Schur functions given by (%
\ref{qdim}), the topological string partition function reads%
\begin{equation}
Z_{\mathrm{top}}^{{\mathbb{X}}_{p}}(q,Q)=\sum_{\lambda }\, (-1)^{p\,
\left\vert \lambda \right\vert}\, Q^{\left\vert \lambda \right\vert}\,
q^{-2n\left( \lambda \right) +\left( p-3\right)\, \kappa(\lambda)/2}\, 
\mathfrak{s}_{\lambda }(q^{i-1/2})^2 \ ,  \label{ZXP}
\end{equation}%
where we have dropped an irrelevant overall constant.

Two distinct simplifications of (\ref{ZXP}) occur for $p=1$%
\begin{equation}
Z_{\mathrm{top}}^{{\mathbb{X}}_{1}}(q,Q)=\sum_{\lambda }\, (-Q)^{\left\vert
\lambda \right\vert}\, q^{-\sum_i\, \lambda _{i}^{2}+\left\vert \lambda
\right\vert }\, \mathfrak{s}_{\lambda }(q^{i-1/2})^2 \ ,
\end{equation}%
and for $p=3$%
\begin{equation}
Z_{\mathrm{top}}^{{\mathbb{X}}_{3}}(q,Q)= \sum_{\lambda }\, (-Q)^{\left\vert
\lambda \right\vert}\, q^{-2n\left( \lambda \right) }\, \mathfrak{s}%
_{\lambda }(q^{i-1/2})^2 \ .
\end{equation}%
As explained in~\cite{Caporaso:2006gk}, here $Z_{\mathrm{top}}^{{\mathbb{X}}%
_{p}}(q,Q)$ is an equivariant partition function which describes a
topological string theory that is generically different from the standard
one. For $p=1$ it gives the usual topological string theory on the resolved
conifold $\mathbb{X}$, and corresponds to a 1-Toda tau-function. For $p=2$
the equivariant topological string theory has partition function $Z_{\mathrm{%
top}}^{{\mathbb{X}}_2}(q,Q)=Z_{\mathrm{top}}^{{\mathbb{X}}_1}(q,Q)^{-1}$,
consistently with the duality (\ref{DTdual}) between the corresponding
Donaldson-Thomas theories. The case $p=3$ corresponds to the local $\mathbb{P%
}^2$ partition function on $\mathcal{O}_{\mathbb{P}^2}(-3)\to \mathbb{P}^2$ 
\cite{Caporaso:2006gk}. In this instance there is no second Casimir term and
hence no quadratic dependence on the boxes of the representation $\lambda_i$%
. This seems to imply that it corresponds to a 2-Toda tau-function, in
contrast to its non-perturbative completion, which is presumably related to
the fact that the local $\mathbb{P}^2$ geometry contains a compact
four-cycle. It would be interesting to better understand in general which
topological string theories correspond to 2-Toda hierarchies, for which the
string theory may be more naturally described using Okounkov's formalism of
double Hurwitz numbers.

A matrix model for the topological string partition function $Z_{\rm
  top}^{\mathbb{X}_p}(q,Q)$ was found to leading orders
in~\cite{Caporaso:2006gk}, and subsequently extended to all orders
using the $q$-deformation of the Plancherel measure by Eynard
in~\cite{Eynard:2008mt}. Explicit connections between Nekrasov
partition functions and topological string amplitudes through the
constructions of matrix models from the point
of view of the Plancherel measure and its generalizations were found
in~\cite{Klemm:2008yu,Sulkowski:2009br,Sulkowski:2009ne}, which
rederive the results of~\cite{Caporaso:2006gk,Eynard:2008mt} from a
more general perspective. These
matrix models were recently extended to more general geometries
in~\cite{Eynard:2010dh}. These matrix models for Nekrasov partition
functions are also related to the infinite chamber limit of the
Donaldson-Thomas matrix models found in~\cite{Sulkowski:2009rw}.

\section{Vicious walkers and stochastic growth representation\label{DTwalk}}

\subsection{Lock-step model}

In \cite{lockstep}, we find an interpretation of certain random unitary
matrix model averages 
\begin{eqnarray}
&& \Big\langle \, \prod\limits_{i=1}^{N}\, \det \big( I+ x _{i}\, U^\dag %
\big)\, \det \big( I+ y _{i}\, U\big)\, \Big\rangle _{U(N)}  \notag \\
&& \hspace{3cm} \ = \ \prod\limits_{j=1}^{N}~\int_{0}^{2\pi }\, \frac{%
\mathrm{d}\theta_j}{2\pi}~ \prod_{i=1}^N\, \left(1+ x_i~{\,\mathrm{e}}\,^{-{%
\,\mathrm{i}\,}\theta_j}\right)\, \left(1+ y_i~{\,\mathrm{e}}\,^{{\,\mathrm{i%
}\,}\theta_j}\right)~\prod\limits_{k<l}\,\big\vert {\,\mathrm{e}}\,^{{\,%
\mathrm{i}\,}\theta _{k}}-{\,\mathrm{e}}\,^{{\,\mathrm{i}\,}\theta _{l}}%
\big\vert ^{2}  \label{average}
\end{eqnarray}
in terms of configurations of weighted non-intersecting lattice paths. The
non-intersecting path model associated to (\ref{average}) is the lock-step
model, with a slight variation. It is explained in detail in \cite{lockstep}
and we summarize their description in what follows. On the $x$-axis, the
allowed points are $x=1,2,\dots,N.$ The procedure is that each point is
moved to the line $y=1,$ according to the rule that each $x$ coordinate must
either stay the same (unit weight) or increase by one (weight $x _{1}$),
always with all $x$ coordinates remaining distinct. This procedure is
repeated a total of $N$ times, with each right diagonal segment at step $j $
weighted by $x _{j}$ as $x$ coordinates are moved to the line $y=j$. Notice
the difference with the original lock-step model, where at each time all
particles move either to their right or to their left with equal
probability. After step $N$, perform another $N$ steps, but now with the
segments either vertical (unit weight) or left diagonal (weight $y _{2N+1-j}$
at step $N+j$). As usual, we have the conditions that the segments do not
intersect and, in addition, are further constrained to return at the line $%
y=2N$ to the same initial $x$ coordinates.

We thus have, as was the case in the Brownian motion description of
Chern-Simons partition functions \cite{dHT}, a reunion condition on the
walkers. The result of~\cite{lockstep} is that the generating function for
this process is given by (\ref{average}). As we have seen in Section~\ref%
{C3matrix}, in the limit $N\rightarrow \infty $ with the weights of the
process taken to be $x _{i}= y _{i}=q^{i-1/2},$ this is the Donaldson-Thomas
partition function $Z_{\mathrm{DT}}^{%
\mathbb{C}
^{3}}\left( q\right) $. Since $N=\infty $ there are infinitely many walkers.
Likewise, the reduced orbifold Donaldson-Thomas partition function $%
\widetilde{Z}_{\mathrm{DT}}^{\,{\mathbb{Y}}}(q,Q)$ is the generating
function for this process with weights $x_i= y_i=\sqrt Q\,(-q)^{i-1/2}$.
More general geometries correspond to higher-dimensional versions of this
process, with the walkers constrained to move independently on coordinate
planes. In this picture, wall-crossing phenomena appear as the creation or
destruction of coordinate planes, and hence particles, and changes in the
weightings of walkers.

The celebrated Robinson-Schensted-Knuth correspondence \cite{Stanley}
establishes a bijection between weighted $N\times N$ non-negative integer
matrices with entries $a_{ij}$ weighted by $\left( x_{i}\, y_{j}\right)
^{a_{ij}}$ and pairs of weighted semi-standard tableaux of content $N$. This
implies that (\ref{SC}) is the generating function for the weighted
matrices, and hence the MacMahon function $M(q)$ is the generating function
for infinite-dimensional matrices with entries $a_{ij}$ weighted by $\left(
q^{(i-1/2)\, (j-1/2)\, a_{ij}}\right) .$ Moreover, in \cite{lockstep} it is
found that the function 
\begin{equation}
{\prod\limits_{i,j=1}^{N}\,\left( 1-x_{i}\, x_{j}\right) ^{-2}}
\end{equation}%
is the generating function for $2N\times 2N$ matrices $[a_{ij}]$ which are
invariant under reflections about the entry $(N+\frac12,N+\frac12)$, i.e. $%
a_{ij}=a_{2N+1-i \ 2N+1-j}$. If we specify $x_{i}=q^{i-1/2}$ and take the
limit $N\rightarrow \infty $, the generating function is the square of the
MacMahon function, a factor that appears in the Donaldson-Thomas partition
functions of the conifold and the $\mathbb{C}^3/\mathbb{Z}_2$ orbifold.

Interestingly enough, instead of considering a symmetry constraint on the
matrices, one can impose constraints on the entries. In particular, if we
constrain them to be $0$ or $1,$ with the entries $a_{ij}$ weighted by $%
\left( x_{i}\, y_{j}\right) ^{a_{ij}},$ then the generating function is the
right-hand side of the dual Cauchy identity (\ref{dualC}) \cite{lockstep}.
Recall that this leads to the Schur expansion (\ref{dual}) of the
Donaldson-Thomas partition function on the resolved conifold. Thus the
reduced partition function $\widetilde{Z}_{\mathrm{DT}}^{\,{\mathbb{X}}%
}\left( q,Q\right) $ can be interpreted as the generating function of such
matrices, with weights $x_{i}=-\sqrt Q\, \left( -q\right) ^{i-1/2}$ and $%
y_{j}=-\sqrt Q\, (-q)^{j-1/2}$. These alternative enumerative descriptions
in terms of infinite non-negative integer matrices are intriguing, in light
of the fact that in general the Donaldson-Thomas partition function $Z_{%
\mathrm{DT}}^{X}$ of a threefold $X$ is a generating function which counts
ideal sheaves on $X$. It would be interesting to formulate a more direct
connection between these infinite-dimensional random matrix theories and the
geometric counting problems.

\subsection{Corner growth model}

It is also possible to relate the Donaldson-Thomas partition functions to
the distributional limit of the corner growth (or last passage) model with
geometric weights \cite{Johansson1}. Let $\omega \left( i,j\right)$, ${%
\left( i,j\right) \in 
\mathbb{Z}
^{2}}$ be independent geometric random variables and define 
\begin{equation}
G(M,N)=\max_{\pi }\, \sum_{(i,j)\in \pi }\, \omega \left( i,j\right) \ ,
\end{equation}%
where the maximum is taken over all up/right paths $\pi$ from $(1,1)$ to $%
(M,N)$; this is called the corner growth model \cite{Johansson1}. It is
shown in \cite{Johansson2} that if $\omega \left( i,j\right) $ are
independent geometric random variables with probability distribution of the
form 
\begin{equation}
\mathcal{P}\big[ \omega \left( j,k\right) =m\big] =(1-x_{j}\, y_{k})\,
(x_{j}\, y_{k})^{m} \ ,  \label{P}
\end{equation}%
then the probability that $G(M,N)$ is smaller than a certain value can be
written in terms of the Schur measure as 
\begin{equation}
\mathcal{P}\big[ G(M,N)\leq t\big] =\Big(\,
\prod\limits_{j,k=1}^{n}(1-x_{j}\, y_{k})\, \Big)~ \sum_{\lambda \, : \,
\lambda _{1}\leq t}\, \mathfrak{s}_{\lambda }\left( x\right) \, \mathfrak{s}%
_{\lambda }\left(y\right) \ .  \label{G(M,N)}
\end{equation}%
Since it is given in terms of the Schur measure, the choice of parameters $%
x_{j}=q^{j-1/2}$ and $y_{k}=q^{k-1/2}$ makes the normalization constant of
the process in the limit $n\rightarrow \infty $ (the normalization constant
for $t\rightarrow \infty $) equal to the Donaldson-Thomas partition function
for $%
\mathbb{C}
^{3}$ in (\ref{DT}).

As in the case of the lock-step model, since the expression (\ref{G(M,N)})
is very general as it involves generic coefficients $x_{j}$ and $y_{k}$, it
also leads to the orbifold partition function (\ref{inv}) by again choosing $%
x_{j}=\sqrt Q\, \left( -q\right) ^{j-1/2}$ and $y_{k}=\sqrt Q\, \left(
-q\right) ^{k-1/2}.$ To describe the conifold partition function (\ref{dual}%
), we need to construct the dual process and hence modify (\ref{P}) to%
\begin{equation}
\mathcal{P}^\vee\big[ \omega\left( j,k\right) =m\big] =(1+x_{j}\,
y_{k})^{-1}\, \Xi_{b}(x_{j}\, y_{k})^{m} \ ,
\end{equation}%
where $\Xi_{b}$ is the endomorphism of the ring of symmetric polynomials
given by $\Xi_b \left(\mathfrak{e}_{\lambda }\right) =\mathfrak{h}_{\lambda
} $ for every Young diagram $\lambda $, with $\mathfrak{e}_{\lambda }$ the
elementary symmetric polynomials and $\mathfrak{h}_{\lambda }$ the
homogeneous symmetric polynomials~\cite{Stanley}. Then 
\begin{equation}
\mathcal{P}^\vee\big[ G(M,N)\leq t\big] =\Big(\, \prod\limits_{j,k=1}^{n}\,
(1+x_{j}\,y_{k})^{-1}\, \Big)~ \sum_{\lambda \, : \, \lambda _{1}\leq t}\, 
\mathfrak{s}_{\lambda }\left(x\right) \, \mathfrak{s}_{\lambda ^{\prime
}}\left(y \right) \ ,
\end{equation}%
and the choice $x_{j}=-\sqrt{Q}\, \left( -q\right) ^{j-1/2}$ and $y_{k}= -%
\sqrt{Q}\, \left( -q\right) ^{k-1/2}$ makes the normalization constant of
the process equal to the partition function (\ref{dual}) in the limit $%
n\rightarrow \infty$. More general geometries correspond to multiple copies
of this process involving independent random variables. Wall-crossing in
this picture is the creation or destruction of independent random variables
and changes of weightings.

A particular case of this stochastic growth model can be understood as a
generalization of the stochastic process underlying the longest increasing
subsequence problem. The distribution of the Poissonized version of the
random variable $L(\alpha)$ describing the longest increasing subsequence in
a random permutation is given by the Gross-Witten model~\cite{BDJ}. The
choice of the distribution (\ref{P}) is a generalization of the original
model introduced in \cite{Johansson2} which has $x_j=x$, $y_k=1$ for all $%
j,k $. If one takes $x=\alpha /N^{2},$ then $G(N,N)$ converges in
distribution to $L(\alpha )$ as $N\rightarrow \infty $, and so one can view $%
G(N,N)$ as a generalization of the random variable $L(\alpha )$.

This model is intimately related to other growth models, as discussed in
detail in \cite{Johansson1,Johansson2}. A growth model is a stochastic
evolution for a height function $h(x,t)$, with $x $ denoting space and $t$
denoting time. An admissible height function has to satisfy $%
h(x+1,t)-h(x,t)=\pm\, 1$ for all $t$. In particular, the discrete
polynuclear growth model is a local random growth model defined inductively
by%
\begin{equation}
h(x,t+1)=\max \big(h(x-1,t)\,,\,h(x,t)\,,\,h(x+1,t)\big)+\omega (x,t+1)
\end{equation}%
with $(x,t)\in 
\mathbb{Z}
\times 
\mathbb{N}
$ and $h(x,0)=0,$ where $\omega (x,t)$ are independent random variables. One
can think of $h(x,t)$ as the height above $x$ at time $t$, so that the map $%
x\mapsto h(x,t)$ describes an interface evolving in time. The special case
where $\omega (x,t)=0$ if $t-x$ is even or if $|x|>t$, and%
\begin{equation}
w(i,j)=\omega (i-j,i+j-1)
\end{equation}%
for $(i,j)\in 
\mathbb{Z}
^{2}$ are independent geometric random variables with probability
distribution (\ref{P}), yields $G(i,j)=h(i-j,i+j-1)$. This growth model is
expected to fall in the Kardar-Parisi-Zhang universality class.

It is also possible to demonstrate the equivalence with other systems, such
as random tilings of Aztec diamonds (which is related to a dimer model), and
non-intersecting walks on a graph~\cite{Johansson1,Johansson2}. The latter
correspondence, in the case of the distribution (\ref{P}) that leads to
Donaldson-Thomas theory, is known in detail \cite{Johanssonreview} but its
description is rather lengthy. It seems that the lock-step model \cite%
{lockstep} described above should be directly related to this version of the
corner growth model since, as we have seen above, the probability (\ref%
{G(M,N)}) without the normalization is the generating function of that
vicious walkers model.

\end{document}